\documentclass[11pt]{article}
\usepackage{xspace,multirow,amsmath,amssymb}
\usepackage{graphicx,hyperref,cite}
\newcommand{\cinst}[2]{$^{\mathrm{#1)}}$~#2\par}
\newcommand{\crefi}[1]{$^{\mathrm{#1)}}$}
\newcommand{\HRule}{\rule{0.4\linewidth}{0.3mm}}

\newcommand{\jpsi}{\ensuremath{{\rm J}/\psi}\xspace}
\newcommand{\pt}{\ensuremath{p_{\rm T}}\xspace}
\newcommand{\xf}{\ensuremath{x_{\rm F}}\xspace}
\newcommand{\QQbar}{\ensuremath{Q\overline{Q}}\xspace}
\newcommand{\qqbar}{\ensuremath{q\overline{q}}\xspace}

\newcommand{\lth}{\ensuremath{\lambda_\vartheta}\xspace}
\newcommand{\lph}{\ensuremath{\lambda_\varphi}\xspace}
\newcommand{\ltp}{\ensuremath{\lambda_{\vartheta\varphi}}\xspace}
\newcommand{\ltilde}{\ensuremath{\tilde{\lambda}}\xspace}

\makeatletter
\newcommand{\lrarrow}{\mathrel{\mathpalette\lrarrow@\relax}}
\newcommand{\lrarrow@}[2]{%
  \vcenter{\hbox{\ooalign{%
    $\m@th#1\mkern6mu\rightarrow$\cr
    \noalign{\vskip2.5pt}
    $\m@th#1\leftarrow\mkern6mu$\cr
  }}}%
}
\makeatother

\begin{document}

\begingroup
\thispagestyle{empty} \baselineskip=14pt
\parskip 0pt plus 5pt

\begin{center}
{\Large\bf \boldmath 
Quarkonium polarization in low-$p_{\rm T}$ hadro-production:\\[3mm]
from past data to future opportunities}

\bigskip\bigskip

Pietro Faccioli\crefi{1}, Ilse Kr\"{a}tschmer\crefi{2} and Carlos Louren\c{c}o\crefi{3}

\bigskip
\textbf{Abstract}

\end{center}

\begingroup
\leftskip=0.4cm \rightskip=0.4cm
\parindent=0.pt

Several fixed-target experiments reported J/$\psi$ and $\Upsilon$ polarization measurements, as functions of Feynman $x$ ($x_{\rm F}$) and transverse momentum ($p_{\rm T}$), in three different polarization frames, using different combinations of beam particles, target nuclei and collision energies. The data form such a diverse and heterogeneous picture that, at first sight, no clear trends can be observed. A more detailed look, however, allows us to discern qualitative physical patterns that inspire and support a simple interpretation: the directly-produced quarkonia result from either gluon-gluon fusion or from quark-antiquark annihilation, with the former mesons being fully longitudinally polarized and the latter being fully transversely polarized. This hypothesis provides a reasonable quantitative description of the J/$\psi$ and $\Upsilon$(1S) polarizations measured in the $x_{\rm F} \lesssim 0.5$ kinematical domain. We provide predictions that can be experimentally tested, using proton and/or pion beams, and show that improved J/$\psi$ and $\psi$(2S) polarization measurements in pion-nucleus collisions can provide significant constraints on the poorly known parton distribution functions of the pion.

\endgroup

\vfill
\begin{flushleft}
\HRule\\

\cinst{1} {LIP, Lisbon, Portugal, Pietro.Faccioli@cern.ch} 
\cinst{2} {ISTA, Klosterneuburg, Austria, Ilse.Kraetschmer@gmx.at}
\cinst{3} {CERN, Geneva, Switzerland, Carlos.Lourenco@cern.ch}

\end{flushleft}
\endgroup

\newpage

\section{Introduction}
\label{sec:intro}

Charmonium and bottomonium production provides an ideal case study for the understanding of hadron formation 
in quantum chromodynamics (QCD)~\cite{Brambilla}. 
Its theoretical description is based on the generally agreed assumption that the charm and beauty quarks 
(the heaviest ones capable of forming bound states)
are heavy enough to allow the factorization of short- and long-distance effects.
Within the non-relativistic QCD (NRQCD) framework~\cite{NRQCD}, in particular, 
perturbative QCD computations provide the production cross sections of the \QQbar pre-resonance
(the ``short-distance coefficients", SDCs),
while the non-perturbative evolution of the \QQbar state to the observed meson 
(the hadronization step) is described by phenomenological parameters
(the ``long-distance matrix elements", LDMEs), determined from fits to experimental data.
Other theoretical approaches have been considered, 
such as the colour-singlet model (CSM)~\cite{Baier:1981uk,Lansberg-HP08} 
and the colour-evaporation model (CEM)~\cite{CEM_BargerPLB,CEM_BargerZPC}.
These theoretical models differ in the choice and classification of the allowed pre-resonance configurations. 
The NRQCD approach foresees the contribution of all possible spin, $S$, 
orbital angular momentum, $L$, total angular momentum, $J$, 
and colour ($c = 1,8$) configurations, $\QQbar(^{2S+1}L_J^{[c]})$, 
organized in an expansion in powers of the relative \QQbar velocity, $v < 1$, 
so that only a small number of leading and sub-leading terms remain quantitatively important.
Instead, the CSM considers that the final-state hadron can only result from a
colour-neutral (singlet) pre-resonance having the same quantum numbers
and the CEM is built upon the assumption that 
one universal hadronization factor per quarkonium state 
(independent of the $S, L, J$ configuration)
multiplies the short-distance \QQbar production cross section.

The fundamental question that all models address is:
how are the observable kinematic properties of the produced quarkonium meson
related to the quantum state of the unobservable \QQbar pre-resonance? 
The answers are different because, among other factors,
the several contributing short-distance processes are scaled by different long-distance weights.
The observable \emph{polarization} of the quarkonium state provides particularly significant information
regarding the hadronization model, 
given that it directly reflects the mixture of $S, L, J$ configurations 
(and polarizations) of the contributing pre-resonance states.
The polarizations of five vector quarkonia
(\jpsi, $\psi$(2S), $\Upsilon$(1S), $\Upsilon$(2S) and $\Upsilon$(3S))
have recently been measured at relatively high transverse momentum, \pt,
both at the Tevatron~\cite{CDF:UpsilonPol}
and at the LHC~\cite{BPH-13-003,BPH-11-023,LHCb_psiPol,LHCb_psi2SPol,LHCb_UpsilonPol}.
These measurements, showing no significant signs of polarization, 
have been addressed in many studies, including analyses based on the 
NRQCD~\cite{Butenschoen:2012px,ChaoPRL,Gong:2012ug,Butenschoen:2012qr,
Faccioli:PLB736,Bodwin:2015iua,Faccioli2018,Faccioli2019} 
and CEM~\cite{iCEM-pol} approaches.

In this paper we devote our attention to low-\pt quarkonium hadro-production,
a kinematical domain complementary to that explored at the LHC.
We start by considering the polarization measurements
reported by several fixed-target experiments, at CERN, DESY and Fermilab,  
using proton or pion beams, in a broad energy range, colliding on targets made of several materials.
The question we address here is: 
can this multitude of low-\pt quarkonium polarization measurements be interpreted in a consistent physical picture?
At first sight, we may think that it is very challenging to see coherent patterns emerging 
from a collection of results obtained in such a diverse set of kinematical conditions, 
affected by several difficulties in the detection and analysis techniques, 
and reported using three different polarization frames.
Nevertheless, a careful look at the experimental results allows us to see that,
while most data points fluctuate around the unpolarized condition, 
there are some tendencies towards strong polarizations in certain kinematical regions.
These qualitative patterns motivate us to consider a simple physical interpretation of low-\pt quarkonium production,
as a superposition of two 2-to-1 processes: gluon-gluon fusion and quark-antiquark annihilation, 
respectively leading to the production of fully longitudinally polarized and fully transversely polarized mesons.
Our study is exclusively focused on the polarization data 
and deliberately follows a model-independent approach. 
Reports on theoretical studies of low-\pt quarkonium cross sections can be found,
for example, in 
Refs.~\cite{Baranov_psi,Baranov_chi,iCEM_MaVogt,iCEM_CheungVogt_psi,iCEM_CheungVogt_upsilon,
Platchkov1,Platchkov2}.

The paper is structured as follows. 
Section~\ref{sec:data} presents and reviews the experimental measurements we have considered.
Section~\ref{sec:indications} discusses possible qualitative indications from the peculiar data patterns,
which are then developed in Section~\ref{sec:model} into a simple model.
Quantitative comparisons between the model and the experimental measurements 
are shown in Sections~\ref{sec:comparison} and~\ref{sec:pions}, 
while predictions for future experiments are provided in Section~\ref{sec:predictions}.

\begin{figure}[p]
\centering
\includegraphics[width=0.72\textwidth]{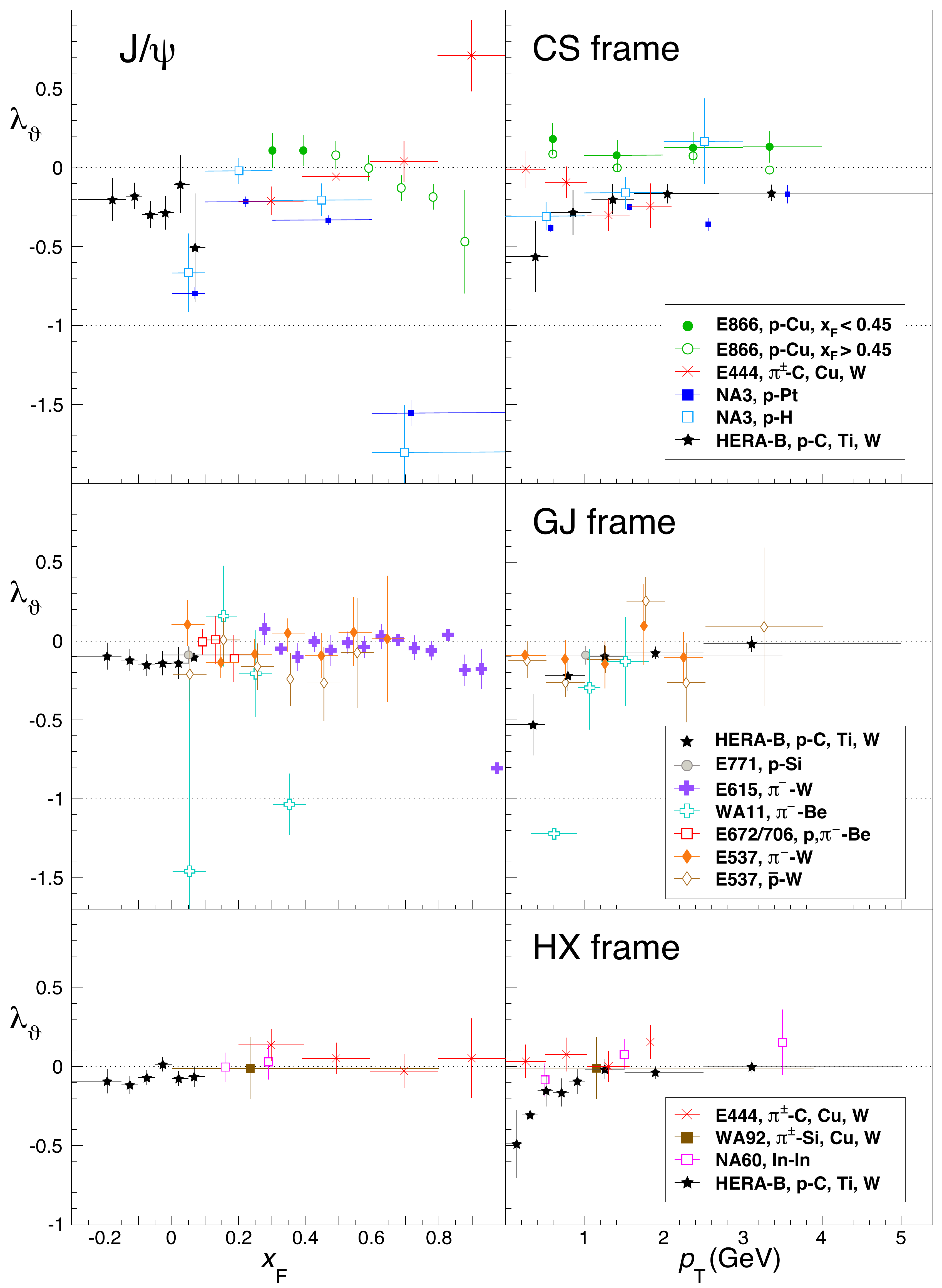}
\caption{The \jpsi polar anisotropy parameter \lth 
measured in the CS, GJ, and HX frames (top to bottom), vs.\ \xf and \pt.}
\label{fig:Jpsi}
\vglue5mm
\centering
\includegraphics[width=0.72\textwidth]{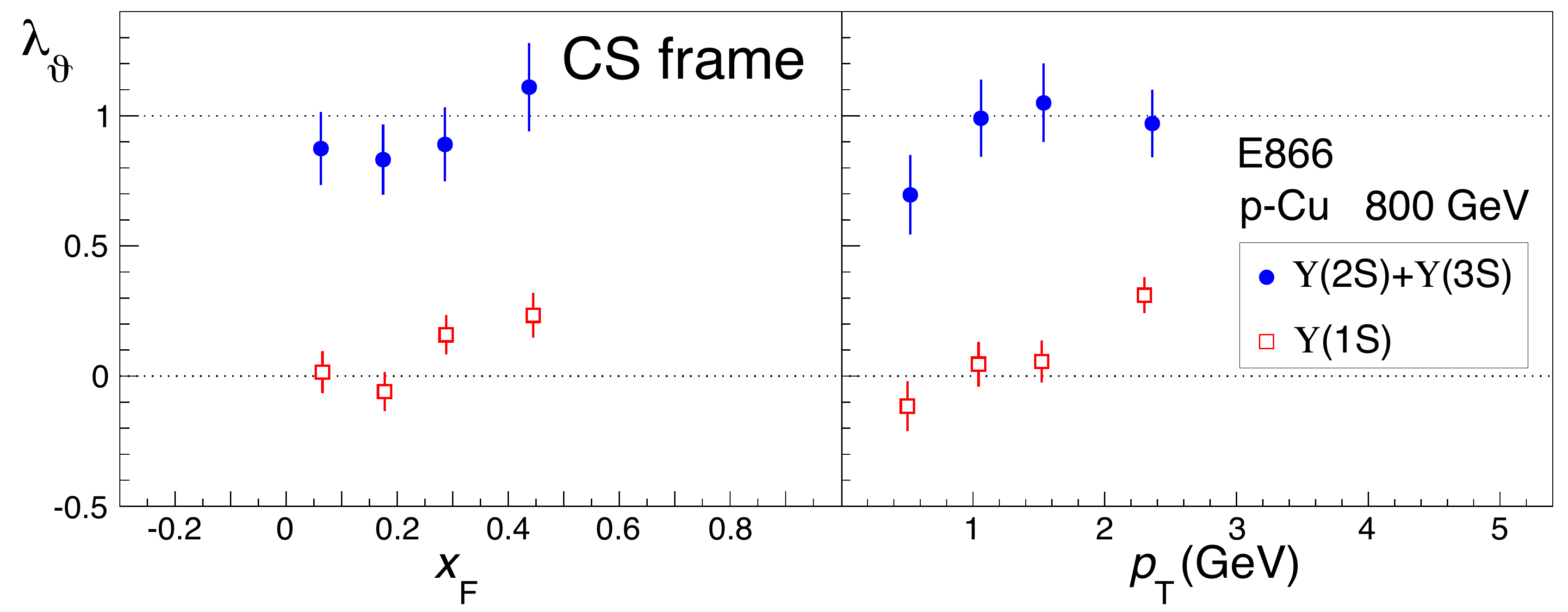}
\caption{The $\Upsilon$(1S) and $\Upsilon$(2S$+$3S) polar anisotropy parameter \lth 
measured by E866 in the CS frame, vs.\ \xf and \pt.}
\label{fig:Upsilon}
\end{figure}

\section{Experimental data}
\label{sec:data}

Figures~\ref{fig:Jpsi} and~\ref{fig:Upsilon} present,
respectively for the \jpsi and $\Upsilon$ states, 
polarization measurements
made by fixed-target experiments, listed in Table~\ref{tab:exps},
using proton or pion beams and several target materials.
The considered observable, shown as functions of \xf and \pt, 
is the polar anisotropy parameter \lth~\cite{FaccioliEPJC2010}.
Most of the measurements address \jpsi 
production~\cite{E537,WA11-PLB,NA60,E444,E615,NA3,WA92,E672-E706-pions,E672-E706-protons,E771,E866-Jpsi,HERA-B}, 
with only one measurement of $\Upsilon$ production~\cite{E866-Upsilon}. 
The ensemble of experiments covers an overall kinematical domain defined by 
$-0.3 \lesssim \xf \lesssim 1$ and $0 < \pt \lesssim 5$\,GeV,
with average \pt between 1.0 and 1.2\,GeV and 
average \pt squared in the $1.5 \lesssim \langle \pt^2 \rangle \lesssim 2.2$\,GeV$^2$ range.

The polarizations were measured in three different frames: 
Collins--Soper (CS)~\cite{CS}, Gottfried--Jackson (GJ)~\cite{GJ} and centre-of-mass helicity (HX), 
where the polarization axis $z$ is defined, respectively, as
the relative direction of the colliding nucleons,
the direction of one of the two nucleons (generally the beam proton),
and the direction of the quarkonium itself with respect to 
the centre-of-mass of the system of the two nucleons.

\begin{table}[t!]
\centering
\caption{\jpsi and $\Upsilon$ polarization measurements in fixed target experiments, 
characterized by several beam energies ($E_{\rm lab}$) and angular coverages,
denoted using \xf, centre-of-mass rapidity ($y_{\rm cms}$) 
or fractional momentum of the beam partons ($x_1$).}
\label{tab:exps}
\renewcommand{\arraystretch}{1.5}
\footnotesize
\begin{tabular}{l c c c c c c c}
\hline
 Exp.\ [Ref.]  & Beam & Target & $E_{\rm lab}$ & $\sqrt{s}$ & $\Delta \xf$ & $\Delta \pt$ & $\langle \pt \rangle$, $\langle \pt^2 \rangle$ \\[-1mm]
                    &                   &            & (GeV)             & (GeV)       &                    & (GeV) & (GeV), (GeV$^2$) \\ 
\hline
\multicolumn{8}{c}{\jpsi} \\
\hline
 E537~\cite{E537} & $\pi^{-}$, $\bar{\rm p}$ & W & 125 & 15.3 & 0.0--0.7 & 0--2.5 & $\langle \pt \rangle = 1.04$ \\
 WA11~\cite{WA11-PLB} & $\pi^{-}$ & Be & 146 & 16.6 & 0.0--0.4 & 0--2.4 & $\langle \pt \rangle = 1.0$ \\
 NA60~\cite{NA60} & In & In & 158 & 17.2 & $y_{\rm cms}$: 0--1 & $\approx$\,0--4 &  \\
 E444~\cite{E444} & $\pi^{\pm}$ & C, Cu, W & 225 & 20.6 & $x_1$: 0.2--1.0 & 0--2.5 & $\langle \pt \rangle = 1.2$ \\
 E615~\cite{E615} & $\pi^{\pm}$ & W & 252 & 21.8 & 0.25--1.0 & 0--5 &  \\
 NA3~\cite{NA3} & $\pi^{-}$ & H, Pt & 280 & 22.9 & 0.0--1.0 & & $\langle \pt^2 \rangle = 1.52, 1.85$ \\
 WA92~\cite{WA92} & $\pi^{-}$ & Si, Cu, W & 350 & 25.6 & $\approx$\,0.0--0.8 & 0--4 &  \\
 E672/706~\cite{E672-E706-pions} & $\pi^{-}$ & Be & 515 & 31.1 & 0.1--0.8 & 0--3.5 & $\langle \pt \rangle = 1.17$ \\
 E672/706~\cite{E672-E706-protons} & p & Be & 530, 800 & 31.5, 38.8 & 0.0--0.6 & & $\langle \pt \rangle = 1.15$, 1.22 \\
 E771~\cite{E771} & p & Si & 800 & 38.8 & $-0.05$--0.25 & 0--3.5 & $\langle \pt^2 \rangle = 1.96$ \\
 E866~\cite{E866-Jpsi} & p & Cu & 800 & 38.8 & $\approx$\,0.0--0.5 & $\approx$\,0--4 &  \\
 HERA-B~\cite{HERA-B} & p & C, Ti, W & 920 & 41.6 & $-0.34$--0.14 & 0--5.4 & $\langle \pt^2 \rangle = 2.2$ \\
\hline
\multicolumn{8}{c}{$\Upsilon$} \\
\hline
 E866~\cite{E866-Upsilon} & p & Cu & 800 & 38.8 & 0.0--0.6 & 0--4 & $\langle \pt \rangle = 1.3$ \\
\hline
\end{tabular}
\end{table}

Since each of the several experiments that measured the polarization of \jpsi mesons 
used different combinations of beam particles, target nuclei and collision energy, 
it is, a priori, not surprising to see that the six panels of Fig.~\ref{fig:Jpsi} 
display a rather scattered overall picture.
The collision energies span a broad range, from $\sqrt{s} = 15.3$ to 41.6\,GeV,
while the target nuclei include eight elements between hydrogen and tungsten.
The beam particles include pions (both charges), protons and antiprotons, 
and even indium nuclei. 
And in the case of secondary beams (e.g., the pion and antiproton cases),
the beam composition is contaminated by some fraction of other particles,
which adds further complexity to the picture.
These complications can be illustrated with a few examples.
E444 collected data with a beam composed of several particles 
($\pi^{\pm}$, $K^{\pm}$, p, $\bar{\rm p}$) 
hitting a target system composed of several materials (C, Cu, W),
the combination $\pi^{-}$-C being the most important.
WA11 collected 40\% of the data at 140\,GeV beam momentum and 60\% at 150\,GeV.
E537 collected data with several beam-target configurations; 
the \jpsi sample is dominated by the $\pi^-$-W combination 
but there is also an important contribution (around 25\% of the events) 
from $\bar{\rm p}$-W collisions,
while the data collected with Be and Cu targets, with both beams, 
is a negligible contamination.

Besides the diversity of collision energies, beam particles and target nuclei,
which surely contributes to 
the visible spread of the data points,
we also need to take into consideration that polarization measurements
are always very challenging and it is quite possible that some of the reported
systematic uncertainties are underestimated (in fact, some of the older results
were even published without mentioning systematic uncertainties).
In particular, most of the measurements were obtained from one-dimensional analyses, 
only considering the $\cos \vartheta$ observable and neglecting acceptance correlations 
between the $\cos \vartheta$ and $\varphi$ variables of the dilepton angular distribution, 
a practice that can easily lead to significantly biased results, 
as discussed in Refs.~\cite{FaccioliEPJC2010,MPLA}.
This might explain why some of the data points shown in Fig.~\ref{fig:Jpsi} 
are outside of the physically allowed range (with $\lth < -1$).

Despite the first impression that the diversity of points form a rather scattered overall picture,
we can see that most of the \jpsi values fluctuate around the $\lth = 0$ limit (unpolarized production),
with some trends towards strong polarizations in certain points of the kinematic domain.
Among all the \jpsi measurements, 
the one published by \mbox{HERA-B}~\cite{HERA-B} stands out
as the only one that considers all three polarization frames (CS, GJ and HX)
and that, furthermore, includes all three shape parameters of the angular distribution 
relevant for parity-conserving decays, \lth, \lph and \ltp~\cite{FaccioliEPJC2010}.
It will, therefore, provide a very useful beacon to guide our extraction
of physically relevant trends (presented in the next section)
from the seemingly cryptic data collection depicted in Fig.~\ref{fig:Jpsi}.

The most salient feature that one can easily see as standing out of the global picture
is the polarization measurement reported by E866 for the 
(unresolved) $\Upsilon$(2S) plus $\Upsilon$(3S) states~\cite{E866-Upsilon}.
While the values reported for the $\Upsilon$(1S) mesons produced with $\xf < 0.45$
cluster around $\lth \sim 0.1$ and are consistent with the \jpsi values
provided by the same experiment, for identical experimental conditions~\cite{E866-Jpsi}, 
those reported for the 2S$+$3S states are surprisingly different: $\lth \sim +1$.
If we exclude the possibility of problems with the experimental measurement, 
this observation reveals an astounding difference 
between the polarizations observed for the excited states and for the ground state.
Given that these three S-wave states are expected to have identical polarizations
when directly produced 
(or when being produced in decays of heavier S-wave states~\cite{BES,CLEO1,CLEO2}),
the observed difference, $\lth{\rm (2S+3S)} - \lth{\rm (1S)} \sim 1$,
seems to be almost impossible to reproduce, 
even resorting to extreme hypotheses
for the polarizations of the $\chi_{bJ}$(nP) states 
and their feed-down contributions to the production of the vector states
(at least in the kinematical conditions of these low \pt measurements).

In the study reported in this paper we followed a cautious approach,
fully developing the model by adapting the (only) free parameter to the \jpsi data, 
without trying to account for the $\Upsilon$ data at all, 
and then comparing the outcome of the computations to the $\Upsilon$ patterns.
Only a posteriori we will discuss what one can infer from that comparison.
Our reluctance in using the remarkable $\lth{\rm (2S+3S)} - \lth{\rm (1S)} \sim 1$ difference
as an \emph{input} of our study is exclusively based on a principle of caution.
%
Polarization measurements are always challenging and this case is
even more demanding because 
the dimuon mass distribution reported by E866 suffers from a poor 
measurement resolution and a daunting signal-to-background ratio, 
so that the (unresolved) 2S and 3S states are not visible as a peak 
on the top of the underlying continuum.
In constrast, 
the $\Upsilon$(1S), $\Upsilon$(2S) and $\Upsilon$(3S) polarizations 
measured by CMS~\cite{BPH-11-023}, 
in much more favourable experimental conditions,
do not show any hint for differences between the three S-wave states.
Clearly, if the patterns shown in Fig.~\ref{fig:Upsilon}
are not disturbed by experimental difficulties, 
their comparison to the CMS results 
points to the importance of the different experimental conditions:
the collision energy, the longitudinal and transverse momentum ranges,
and the use of a nuclear target (Cu) in the E866 data.
In any case, it would certainly be very useful to have the fixed-target 
measurement repeated by another experiment,
with improved detection capabilities.

\section{Overall qualitative physical indications}
\label{sec:indications}

The model presented in this paper, 
described in much more detail in the next section and 
then quantitatively tested in Section~\ref{sec:comparison},
can be very briefly summarised by saying that
low \pt quarkonium production is dominated by two processes,
the quarkonia produced in gluon-gluon fusion having longitudinal polarization 
and those produced in quark-antiquark annihilation having transverse polarization.
The model is inspired by two qualitative physical observations revealed by 
a careful look at the \jpsi polarization patterns shown in the six panels of Fig.~\ref{fig:Jpsi}.

\begin{figure}[t!]
\centering
\includegraphics[width=0.6\textwidth]{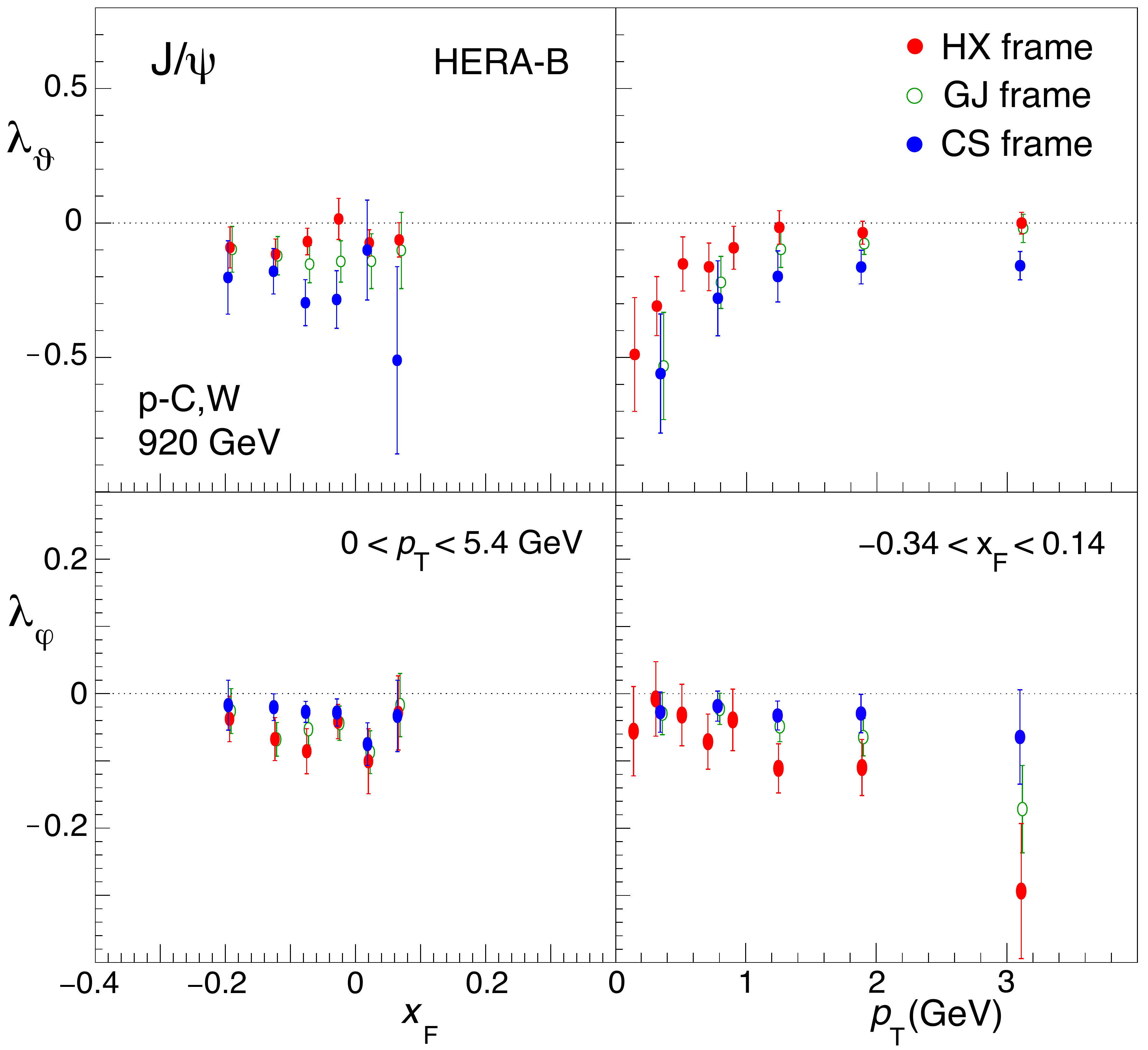}
\caption{The \jpsi \lth and \lph parameters measured by \mbox{HERA-B} 
in the CS, GJ and HX frames, vs.\ \xf and \pt.}
\label{fig:hierarchy}
\end{figure}

The first observation follows from the \mbox{HERA-B} measurements~\cite{HERA-B}, 
which were performed considering in turn all three polarization frames and
provide all three shape parameters, \lth, \lph and \ltp.
As is already suggested by the global picture of \jpsi measurements 
(presented in Fig.~\ref{fig:Jpsi}) 
and can be more clearly and precisely seen in the top panels of Fig.~\ref{fig:hierarchy}, 
the magnitude of the polar anisotropy parameter \lth is systematically larger in the CS frame 
and smaller in the HX frame. 
At the same time, as shown in the bottom panels of Fig.~\ref{fig:hierarchy},
the azimuthal anisotropy parameter \lph increases in magnitude 
following the inverse hierarchy, being the largest in the HX frame and the smallest in the CS frame.
It is important to appreciate that the differences between the values 
measured in the three frames \emph{are} significant, 
contrary to what the displayed uncertainties could indicate at first sight, 
because the three sets of data points were obtained using exactly the same events and,
hence, their uncertainties are strongly correlated.
In other words, the \emph{differences} between the three sets of parameters 
have much smaller uncertainties than those represented by the error bars,
so that we can say that these differences are significantly larger than zero.

These two observed hierarchies reflect, to start with, 
the geometrical difference between the three frame definitions: 
the GJ polarization axis has always, for every \pt and \xf values, 
an intermediate direction between the CS and HX ones. 
On the other hand, and more importantly, the direction of the hierarchy, 
with the CS frame showing the largest polarization effect with the smallest azimuthal anisotropy, 
has one clear physical interpretation: 
the CS axis is the one that most naturally reflects the alignment of the \jpsi angular momentum. 
This provides relevant information regarding the topological nature of the involved processes: 
they must be, predominantly, of the 2-to-1 kind 
($h_1 \, h_2 \to \jpsi$),
where the produced object directly inherits the angular momentum state 
of the system of colliding partons, 
which is polarized along the direction of the collision. 
One can expect, a priori, that most of the mesons produced in 2-to-2 processes 
($h_1 \, h_2 \to \jpsi + X$, where the \jpsi is emitted with a recoil hadron)
have \pt much larger than the intrinsic momenta of the colliding partons, 
so that they have a negligible contribution to the low \pt yield with respect to the 2-to-1 processes. 
The observed polarization hierarchy further disfavours 2-to-2 contributions.
In fact, in 2-to-2 processes the polarization legacy of the partons 
is shared between the two final objects and, 
while the angular momentum balance is more complex 
and depends on the coupling of the final states to the intermediate virtual particles,
one can say that, in general, 
a natural alignment along the direction of the colliding partons is excluded.

An interesting illustration is provided by the case of the individual processes contributing to 
Drell--Yan production (DY)~\cite{FaccioliPRD83}.
At the lowest order, DY production is a 2-to-1 process, 
characterized by a ``natural" polarization along the collision direction, 
approximated by the CS axis. 
On the other hand, 2-to-2 processes ($t$- or $s$-channel) 
naturally lead to polarizations along the GJ and HX axes.

In the case of quarkonium production, 
the final state (the experimentally observable quarkonium state) 
evolves from an intermediate (singlet or octet) \QQbar pre-resonance state,
through the possible emission of one or more soft gluons.
But the emission of one soft gluon (or more) in the bound state formation process
does not qualify the process as ``2-to-2", 
given that we are talking about the process that produces the \QQbar state:
the emitted gluon is not to be seen as a final-state object of a 2-to-2 topology.
Indeed, as long as the mass difference between the real final state 
and the virtual intermediate state 
is smaller than the total momentum ($\sqrt{(s/4) \, \xf^2 + \pt^2}$) 
of the observed state in the centre-of-mass of the colliding nucleons, 
the natural polarization direction of the final state coincides with the one of the intermediate state 
(that is, the direction of the colliding partons, in 2-to-1 processes) 
as discussed in Ref.~\cite{chicPol} for the case of the radiative $\chi_c$ decays to \jpsi.
These considerations motivate our assumption that the 2-to-1 $\qqbar \to \QQbar$ 
and $gg \to \QQbar$ scattering processes dominate, leading to strongly polarized quarkonia.

\begin{figure}[t]
\centering
\includegraphics[width=0.4\textwidth]{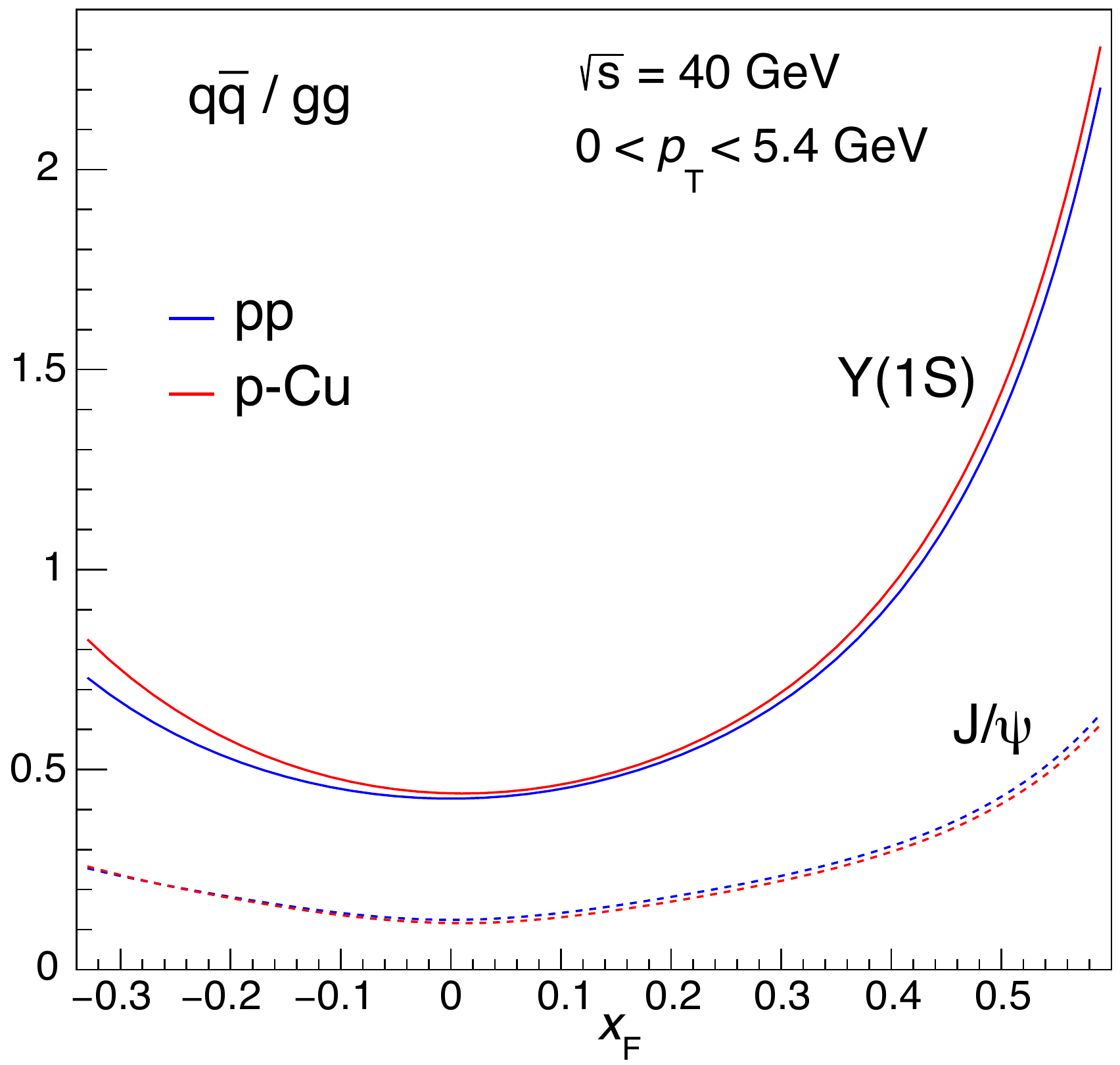}
\caption{The \qqbar to $gg$ parton luminosity ratios for \jpsi and $\Upsilon$(1S) production, 
vs.\ \xf, illustrated for conditions similar to those of the E866 and \mbox{HERA-B} experiments.}
\label{fig:PDFratios}
\end{figure}

The second experimental indication is the observation of trends towards 
longitudinal \jpsi polarization at small $|\xf|$ (``mid-rapidity").
%
It is reasonable to suppose that this behaviour might be correlated with the
relative dominance of gluon-gluon fusion at mid-rapidity, 
the \qqbar annihilation process becoming more relevant 
as we move away from that central region.
This hypothetical correlation can be better appreciated by looking at the ratio 
between the \qqbar and the $gg$ ``luminosities'', both computed as the 
product of the corresponding parton distribution functions (PDFs).
Figure~\ref{fig:PDFratios} shows the \xf dependence of this $\qqbar/gg$ ratio, 
computed for $\sqrt{s} = 40$\,GeV and for the \jpsi and $\Upsilon$(1S) cases, 
using the CT14NLO~\cite{CT14NLO} set of proton PDFs, 
as provided by the LHAPDF library~\cite{LHAPDF}.
The p-Cu curves were computed with the target PDFs (both for the protons and for the neutrons)
modified to reflect nuclear effects, as provided by the EPPS16 package~\cite{EPPS16}.
For the collision energies relevant for the E866 and \mbox{HERA-B} experiments,
the nuclear effects on the PDFs play a negligible role.
We clearly see that, as expected, the ratio has a minimum in the $\xf \sim 0$ region
and increases as we move away from mid-rapidity.

It is also clear that
the ratio is considerably higher for the $\Upsilon$(1S) than for the \jpsi, 
reflecting the fact that, relatively speaking, 
\qqbar annihilation becomes more important in the production of heavier particles.
This  is another important piece of evidence supporting the hypothesis that
\qqbar annihilation leads to transversely polarized quarkonia, 
given that, overall, the measurements collected in Figs.~\ref{fig:Jpsi} and~\ref{fig:Upsilon}
indicate that the observed polarizations are more transverse for the $\Upsilon$ states 
than for the \jpsi.

This correlation between the measured quarkonium polarization patterns 
and the relative importance of the computed parton-parton luminosities 
is the central motivation of our postulate 
(to be further examined by confronting its implications with the existing measurements)
that $gg$ fusion produces longitudinally polarized mesons while
those produced through \qqbar annihilation are transversely polarized.
A physical justification supporting this assumption is beyond the scope of this paper,
where we take a data-driven approach and let the measured patterns be our guiding principles.
Nevertheless, it is not difficult to conceive that \qqbar-induced \jpsi and $\Upsilon$ 
production might be analogous to the DY case and that, therefore, 
those quarkonia should be transversely polarized 
(angular momentum projection $J_z = \pm 1$) 
because of helicity conservation in the coupling between the annihilating quarks and a gluon. 
The longitudinal polarization of the quarkonia produced in $gg$ fusion, 
on the other hand, may be justified, for example, 
 with a dominating $J_z = 0$ projection of the $gg$ system 
(then inherited by the \QQbar state), 
which would be a necessary condition, forced by angular momentum conservation, 
if the scattering gluons were transversely polarized and formed a $J = 1$ state. 
Naturally, the CEM and NRQCD approaches are not expected to accommodate with the same proportion 
these two production channels 
(nor, more generally, any pair of oppositely-polarized processes), 
given the (in principle different) foreseen contributions of intermediate virtual \QQbar states 
of different angular momentum quantum numbers. 
In fact, the scenario where $gg$ fusion and \qqbar annihilation produce 
quarkonia with maximally-different polarizations is the one most propitious 
for a test of the hadronization model.

The picture is made more complex, however, by the significant contribution of 
feed-down decays from P-wave states.
Let us consider first the feed-down from an excited \emph{vector} quarkonium. 
As long as mother and daughter have the same mechanism of production from partonic scattering, 
the feed-down decays from heavier vector states are ``invisible" from the polarization point of view. 
This is confirmed by the observation that the \jpsi mesons produced in $\psi(2\mathrm{S}) \to \jpsi \, \pi \pi$ decays 
have the same polarization as the $\psi$(2S) mesons themselves~\cite{BES}
and by the analogous observation made in the $\Upsilon$ family~\cite{CLEO1,CLEO2}.
On the contrary, 
P-wave states have, in general, different production mechanisms with respect to the vector states.
Moreover, the $\chi_c$ and $\chi_b$ mesons decay to the \jpsi and $\Upsilon$ mesons
with the emission of a transversely polarized photon, 
which alters the spin-alignment of the \QQbar~\cite{chicPol}.
Therefore, we should expect that
the \jpsi and $\Upsilon$ mesons produced in decays of P-wave states 
have different polarizations with respect to the directly produced ones. 
In particular, 
if large fractions of the observed vector quarkonia are produced through $\chi$ feed-down decays,
we will probably observe a strong reduction of the transverse or longitudinal polarizations 
that could be measured if direct production were the dominating mechanism.

The \jpsi and $\Upsilon$ feed-down fractions from $\chi$ mesons depend on the experimental conditions: 
for example, they can be different if $gg$ fusion or \qqbar annihilation individually dominate, 
since different selection rules between the initial state and the final S- and P-wave states are expected in the two cases. 
We will consider a ``central'' scenario and two ``extreme'' ones, 
characterized by different values of the feed-down fractions; 
together, these scenarios should represent a reasonably conservative ``uncertainty margin", 
likely to cover the real values of these input variables in our model.
Our central scenario assumes that the observed \jpsi sample is affected by 
a feed-down fraction from $\chi_c$ decays of around 19\%, 
for both $gg$ and \qqbar production, 
corresponding to the central value of the \mbox{HERA-B} measurement~\cite{HERAB_chic}. 
The two extreme scenarios assume that the $\chi_c$ feed-down has 
a maximal impact on the observable prompt \jpsi polarization, as will be specified in the next section.
Therefore, for these scenarios we use the larger value of 25\%, 
representing an upper limit derived taking into account also 
the CDF~\cite{CDF_chic} and LHCb~\cite{LHCb_chic} measurements.
The feed-down fractions in the bottomonium family are not well known, 
especially in the low-\pt range relevant for the fixed-target results 
that we are addressing in this paper.
On the basis of LHCb measurements at forward rapidity~\cite{LHCb_chib} 
and of extrapolated trends of mid-rapidity 
LHC cross sections~\cite{Faccioli2018b}, 
we will assume that, for the central scenario, the $\Upsilon$(1S) and $\Upsilon$(2S+3S) results of E866
are affected by $\chi_b$ feed-down contributions of around 45\% and 25\%, 
respectively;
for the two extreme scenarios we will use the respective upper limits of 60\% and 50\%.

Before concluding this section, 
it is important to note that our study is exclusively devoted to 
quarkonium production in the $\xf < 0.5$ domain
and does not address the high \xf region,
where a trend towards longitudinal polarization has been seen
by E615 and E866 (in pion- and proton-nucleus collisions, respectively).
Although certainly interesting in their own right~\cite{Berger1979,Berger1980}, 
this edge of phase space is likely to be dominated by 
processes that are not covered by the model that we discuss in this paper.

\section{Description of the model}
\label{sec:model}

As anticipated in the previous section, 
our model is based on two main assumptions:
1)~\QQbar production is dominated by 2-to-1 topologies
(qg contributions, producing at least one additional object besides the quarkonium, 
are, therefore, considered negligible);
2)~the $gg \to \QQbar$ and $\qqbar \to \QQbar$ processes lead, 
respectively, to fully longitudinal and fully transverse polarizations 
of the \emph{directly produced} \jpsi, $\psi$(2S) and $\Upsilon$ mesons. 
For these ``natural'' polarizations 
we assume as quantization axis the (unobservable) relative direction of the colliding partons, 
which does not coincide (event-by-event) with the CS axis 
because of the (small but) nonzero parton transverse momenta, 
$\vec{k}_{1\rm{T}}$ and $\vec{k}_{2\rm{T}}$.

In fact, for measurements performed at low \pt, small $| x_\mathrm{F} |$ and 
light particles (the \jpsi in our case),
the transverse component of the parton motion inside the colliding hadrons 
has an effect on the observable polarization.
For the scope of the present discussion, 
the meaning of $\vec{k}_{\mathrm{T}}$ is extended with respect to 
the bare intrinsic momentum owned by partons 
for being confined inside a hadron of finite dimensions 
($\Delta p \sim 1/(1\,\mathrm{fm}) = 0.2$\,GeV), 
also considering other effects occurring during the scattering process 
and possibly influencing the direction of the partonic collision 
(soft gluon emissions, multiple scattering in the nuclear target, etc.). 
We will assume that, given these extra sources of transverse momentum kick, 
the parton $k_{\mathrm{T}}$ reaches a magnitude
$\langle k_{\mathrm{T}}^2 \rangle = \mathcal{O}(1\,\mathrm{GeV}^2)$, 
compatible with the measured \pt distributions: 
$\langle\pt^2\rangle \simeq 2 \, \langle k_{\mathrm{T}}^2 \rangle 
\simeq 2\,\mathrm{GeV}^2$ 
(see, e.g., Ref.~\cite{HERAB_kinematics}).
The parton transverse momenta also provide the 
only source of quarkonium \pt considered in the model.

\begin{figure}[t]
\centering
\includegraphics[width=0.6\linewidth]{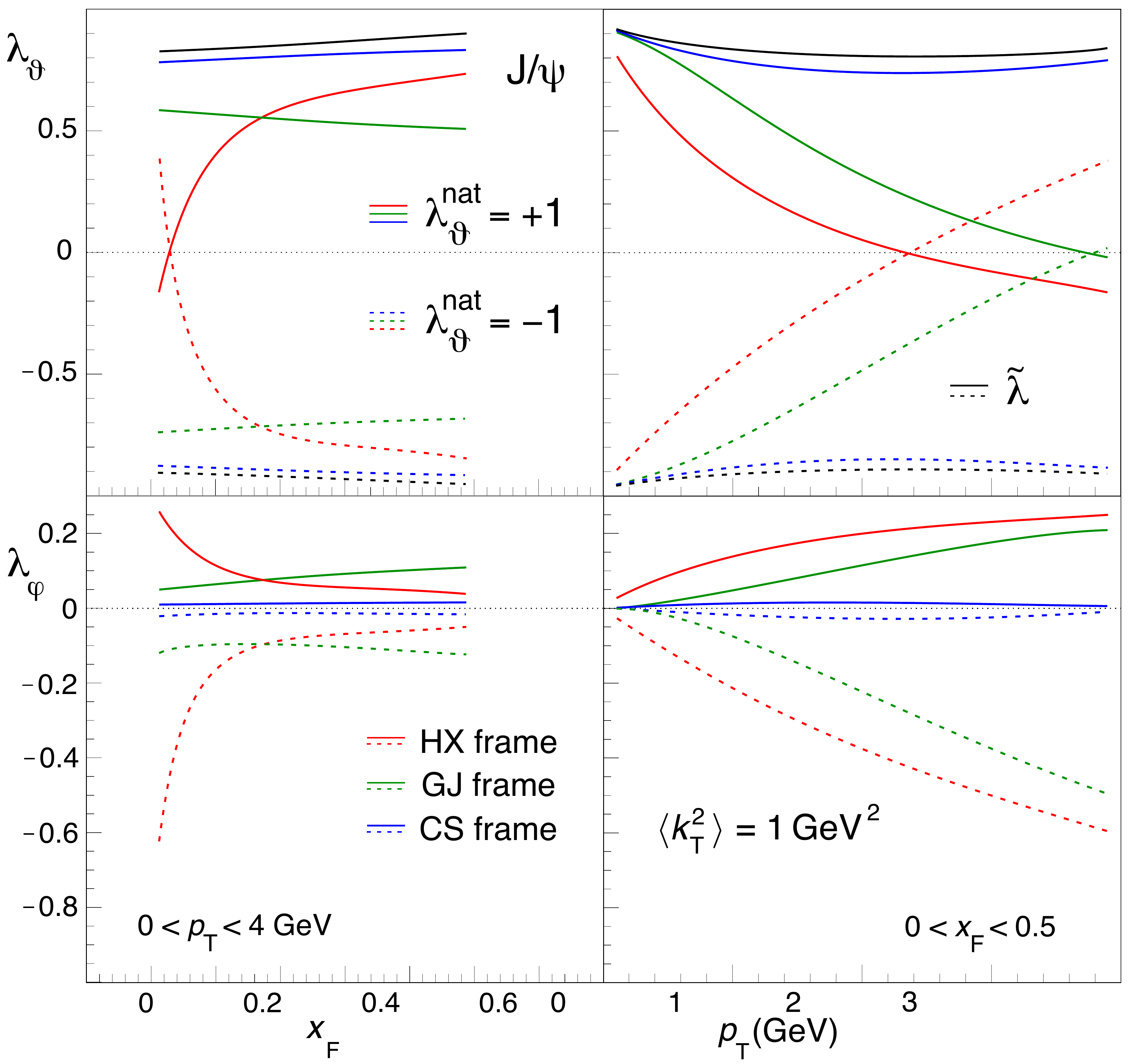}
\caption{The parameters \lth and \lph as would be observed in the CS, GJ and HX frames 
for fully transverse and longitudinal \jpsi polarizations 
along the direction of the colliding partons, 
in the \mbox{HERA-B} experimental conditions, 
for $\langle k_{\rm{T}}^2 \rangle = 1\,\mathrm{GeV}^2$. 
The angular distribution parameters are shown as a function of \xf averaged over \pt (left)
and vice versa (right). 
The invariant polarization parameter \ltilde, 
shown in gray, is by definition identical in the three observation frames.}
\label{fig:kTeffect}
\end{figure}

The vectors $\vec{k}_{1\rm{T}}$ and $\vec{k}_{2\rm{T}}$ are generated in space, 
with the two moduli following a Gaussian distribution 
of variance $\langle k_{\rm{T}}^2 \rangle = 1\,\mathrm{GeV}^2$ 
and the azimuthal angles $\phi_1$ and $\phi_2$ following uniform distributions. 
While the $\vec{k}_{\mathrm{T}}$ effect has a negligible influence for $\Upsilon$ production,
it has a significant impact in the observable \jpsi polarization,
as illustrated in Fig.~\ref{fig:kTeffect} 
for fully transverse (solid lines) and fully longitudinal (dashed lines) natural polarizations, 
in the conditions of the \mbox{HERA-B} experiment but in a slightly larger (and positive) \xf range. 
The main effect (ignoring the slight \pt- and \xf-dependent modulations)
is that the magnitude of the \lth parameter measured in the CS frame
is reduced with respect to the values generated in the parton-parton frame,
by about 20\% and 10\% for the fully transverse and fully longitudinal polarizations, 
respectively.

As expected, the CS frame remains the best approximation of the natural frame, 
because the \lth values observable in the GJ and HX frames 
depart more significantly from the natural one, an effect increasing with \pt. 
The \lph values seen in the three observation frames are also shown in 
Fig.~\ref{fig:kTeffect}, in the bottom panels,
with the expected inverted hierarchy, similar to the one seen by \mbox{HERA-B}, 
as previously pointed out. 
There is a nonzero \lph value also in the CS frame, 
as a result of the rotation from the natural frame. 
However, the transformation from the natural parton-parton to the ``laboratory" CS frame 
is not a simple rotation in the production plane (around the $y$ axis), 
as is the case between any two of the three experimental frames. 
While the magnitude of the polar anisotropy decreases exactly as in such ordinary frame rotations
(for the same rotation angle), 
the correspondingly arising azimuthal anisotropy, $\lph \ne 0$,
no longer geometrically compensates the $| \lth^{\rm CS} |$ decrease. 
In fact, in this case the rotation plane 
(formed by the parton-parton and proton-nucleon relative momentum directions) 
does not coincide with the experimentally defined production plane. 
The angle between the two planes changes from one event to the next, 
so that the azimuthal anisotropy resulting from the tilt between the natural polarization axis 
and the experimental axis tends to be smeared out in the integration over all events.
Consequently, the invariant polarization parameter \ltilde~\cite{Faccioli:2010ej,bib:Faccioli-PRD-FrameInv}
(shown as gray curves on the top panels of Fig.~\ref{fig:kTeffect}), 
while slightly closer to the natural value, does not return its full magnitude: 
the natural polarization is unrecoverably smeared in any observation frame.

The relative proportion of quarkonia \emph{directly} produced through 
\qqbar annihilation and $gg$ fusion processes and, therefore, 
their observable mixture of longitudinal and transverse polarizations,
is fully determined in the model by the product of two ratios, $R$ and $r$.
The first one is the ratio between the \qqbar and $gg$ parton densities,
\begin{equation}
R = \frac{ \sum_q [ F_1^{q}(x_1,\hat{s}) \, F_2^{\overline{q}}(x_2,\hat{s}) 
+ F_1^{\overline{q}}(x_1,\hat{s}) \, F_2^{q}(x_2,\hat{s}) ] }
{ F_1^{g}(x_1,\hat{s}) \, F_2^{g}(x_2,\hat{s}) } \, , 
\label{eq:R_parton_densities}
\end{equation}
where $\hat{s} = M_{\mathcal{Q}}^2 + \pt^2$ and $x_1 \, x_2 = \hat{s} / s$,
with $M_{\mathcal{Q}}$ being the quarkonium mass and 
$\pt = |\vec{k}_{1\rm{T}} + \vec{k}_{2\rm{T}}|$; 
the indexes 1 and~2 refer, respectively, to the beam proton and the target nucleon, 
and the sum is made over the participating quark flavours ($q=u,d$). 
Nuclear modification factors for the nucleon in the target, 
different for sea and valence quarks, 
are computed with the EPPS16 model~\cite{EPPS16} and
applied in the definition of $F_2$.

The second ratio is the one between the \qqbar and $gg$ partonic cross sections,
\begin{equation}
r = \frac{ \hat{\sigma}(\qqbar \to \mathcal{Q}) }{ \hat{\sigma}(gg \to \mathcal{Q}) } \, ,
\label{eq:R_parton_xsections}
\end{equation}
assumed to be universal, that is, 
identical for all considered vector quarkonia, $\mathcal{Q} = \jpsi$, 
$\psi{\rm (2S)}$, $\Upsilon$(nS). 
%
%
In principle, one might be able to evaluate $r$ within the context of specific 
model-dependent approaches, such as, for example, the NRQCD framework.
It should be noted, however, that $r$ is the ratio of the partonic cross sections,
depending not only on the ``short-distance parton-level cross sections" (the SDCs),
which can be computed in perturbative QCD, but also on the probabilities 
of the transitions from the \QQbar ``pre-resonances" (singlet and octet states)
into the final quarkonium state (the LDMEs).
These probabilities represent non-perturbative evolution processes and
are presently not calculated, 
but rather determined from global analyses of collider data.
%
Besides, they are a priori different in the \qqbar and gg cases, which, in general, 
produce pre-resonances of different angular momentum properties.
In our study we deliberately try to remain as agnostic as possible regarding 
model-dependent inputs, so that we treat $r$ as an empirical parameter, 
adjusted through the analysis of the \jpsi data.

The resulting natural polarization parameter $\lambda$ (in the parton-parton CS frame), 
for a given mixture of \qqbar and gg events (expressed by $R \times r$),
is determined according to the sum rule presented in Eq.~11 of Ref.~\cite{FaccioliEPJC2010},
reported here as a function of the \qqbar and gg fractions,
$f_{\qqbar} = R \times r / (1 + R \times r)$ and $f_{gg} = 1 / (1 + R \times r)$,
and of the corresponding assumed polarizations, 
$\lambda^{\qqbar}$ and $\lambda^{gg}$:
\begin{equation}
\lambda = \frac{ f_{\qqbar} \, \lambda^{\qqbar} / (3 + \lambda_\vartheta^{\qqbar} ) 
                      + f_{gg} \, \lambda^{gg} / (3 + \lambda_\vartheta^{gg} ) }
                       { f_{\qqbar} \, / (3 + \lambda_\vartheta^{\qqbar} ) 
                      + f_{gg} \, / (3 + \lambda_\vartheta^{gg} ) }
\end{equation}
This expression is explicitly \xf dependent because of the presence of $R$ 
(that is, of the PDFs) in the \qqbar and gg fractions, 
while a further kinematic dependence, also on \pt, 
is acquired by the polarization parameters when translated to the observable frames (CS and HX);
this translation is performed by generating pseudo-events with a Monte Carlo method.


To turn the polarizations determined in this way, 
for the directly-produced quarkonium states, 
into values that can be compared with the measured data,
we need to take into account the effect of the 
unknown feed-down contributions from the $\chi_c$ or $\chi_b$ states. 
We do so by considering three alternative scenarios,
two of them representing extreme hypotheses.
These scenarios have two kinds of ingredients: 
the natural \lth values for the \jpsi or $\Upsilon$ mesons
produced in the decays of the $\chi_1$ and $\chi_2$ states, and the feed-down fractions from each of those states.

Concerning the polarizations, 
the central-value hypothesis corresponds to assuming that
a)~\qqbar production leads to a (vector or P-wave) 
quarkonium state with angular momentum projection $\pm 1$, that is
(besides the already mentioned $\lth = +1$ for the directly produced vector quarkonia),
$\lth = -1/3$ for \jpsi or $\Upsilon$ mesons from $\chi_1$ or $\chi_2$ decays~\cite{chicPol}; 
and that
b)~for $gg$ production the assumed angular momentum projection is 0, 
meaning $\lth = -1$, $+1$ and $-3/5$, 
respectively for directly produced \jpsi or $\Upsilon$ mesons,
and for those coming from $\chi_1$ and $\chi_2$ decays~\cite{chicPol}.
As variations of these hypotheses, defining the two alternative scenarios,
we consider the extremes of the physical intervals for the polarizations of \jpsi or $\Upsilon$ mesons
from $\chi_1$ and $\chi_2$ decays, $[-1/3, +1]$ and $[-3/5, +1]$, respectively. 
The scenarios using the most longitudinal and transverse $\chi$ polarizations
are referred to by the labels ``lower'' and ``upper'', respectively.
%
For clarity, the natural polarizations assumed in the three scenarios are summarized in Table~\ref{tab:lambdas}.

\begin{table}[ht]
\centering
\caption{Values of \lth considered in the generation of the \jpsi or $\Upsilon$ mesons 
resulting from feed-down decays of $\chi_1$ or $\chi_2$ mesons produced through
$gg$ fusion or \qqbar annihilation, in the baseline ``central'' scenario and in two 
extreme scenarios, ``lower'' and ``upper'', leading, respectively, 
to the most longitudinal and most transverse values for the natural polarization of the total prompt-\jpsi polarization.
In all scenarios, the directly-produced vector states are generated 
with $\lth = -1$ and $+1$ for $gg$ fusion and \qqbar annihilation, respectively.}
\label{tab:lambdas}
\renewcommand{\arraystretch}{1.4}
\begin{tabular}{c c c c}
\hline
 & & $\lth^{\chi_1}$ & $\lth^{\chi_2}$\\
\hline
\multirow{2}{*}{central} & $gg$ & $+1$ & $-3/5$ \\
                                   & \qqbar & $-1/3$ & $-1/3$ \\
\hline
lower & $gg$, \qqbar & $-1/3$ & $-3/5$ \\
upper & $gg$, \qqbar & $+1$ & $+1$ \\
\hline
\end{tabular}
\end{table}

As already mentioned in the previous section, 
the small mass difference between the mother and daughter particles, 
in all considered cases, 
ensures that the natural angular momentum alignment direction is preserved in the decay~\cite{chicPol}:
also for indirect production we use, therefore, the parton-parton direction as quantization axis.

For the total $\chi$ feed-down fraction, $R_{\chi_1} + R_{\chi_2}$, 
we assume, as mentioned in the previous section, 
the values 19\%, 45\% and 25\% for, respectively, the \jpsi, the $\Upsilon$(1S) and the $\Upsilon$(2S+3S) cases 
in the central scenario; the corresponding values in the extreme scenarios are 25\%, 60\% and 50\%.
We assume the ratio between the contributions of the two states to be $R_{\chi_1} / R_{\chi_2} = 1$ in the three scenarios, 
after verifying that its variation within the \mbox{0.6--1.4} range established by \mbox{HERA-B}~\cite{HERAB_chic} 
does not lead to significant changes in the results.
Actually, the range of hypotheses assumed for the $\chi$ feed-down should be wide enough 
to cover possible dependences of the inputs on the experimental conditions, 
such as the \xf and \pt ranges of the different measurements.

\section{Data vs.\ model for p-nucleus collisions}
\label{sec:comparison}

\begin{figure}[t!]
\centering
\includegraphics[width=0.89\linewidth]{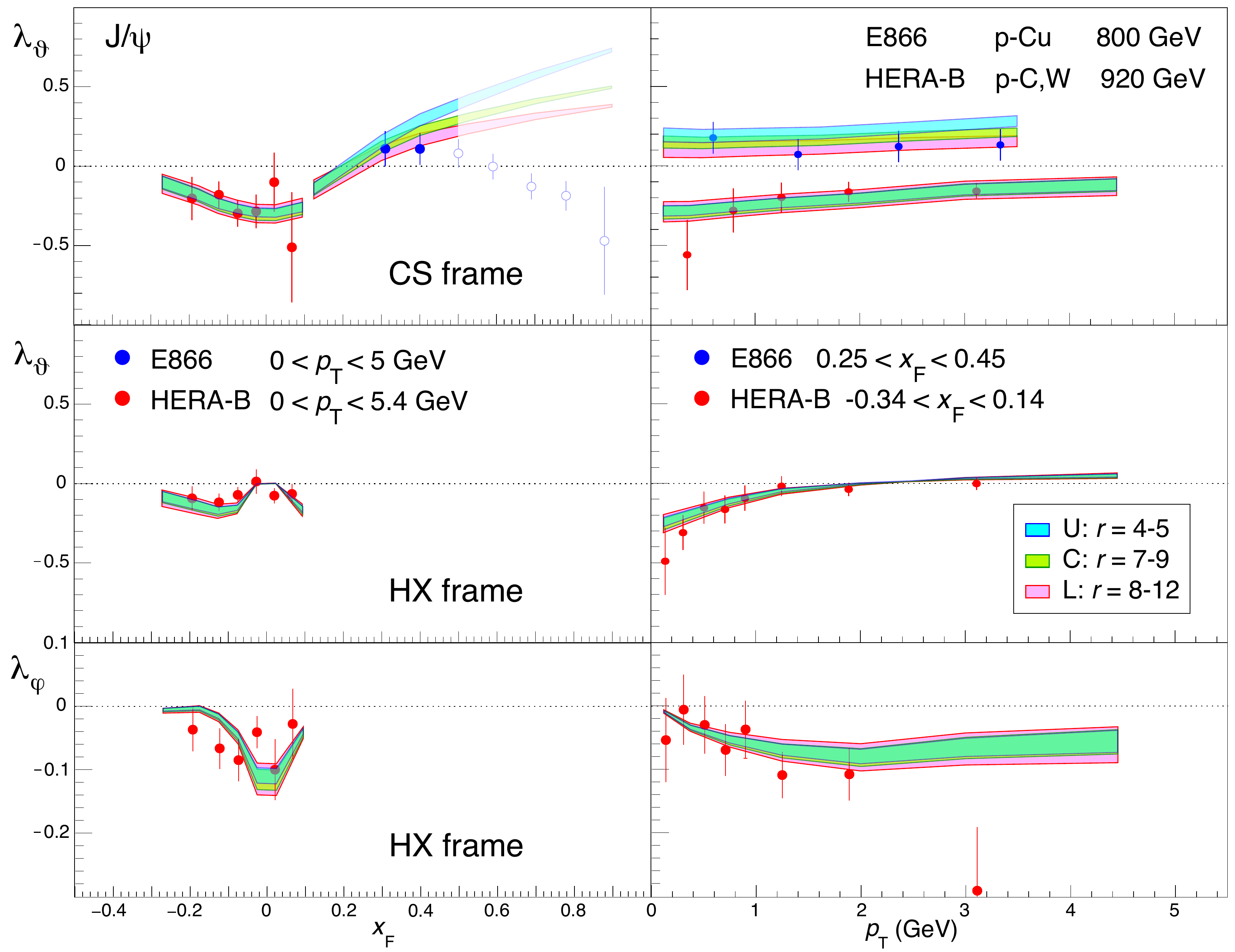}
\caption{The \xf (left) and \pt (right) dependences 
of the \jpsi polarization parameters 
\lth in the CS (top) and HX (middle) frames,
and \lph in the HX frame (bottom),
as measured by \mbox{HERA-B} (red points) and E866 (blue points).
The cyan, green and magenta bands represent, respectively, 
the upper (U), central (C) and lower (L) scenarios described in the text.
Since our model is not expected to describe high-\xf quarkonium production,
the E866 measurements for $\xf > 0.45$ are depicted as open circles.}
\label{fig:results_Jpsi}
\end{figure}

Figure~\ref{fig:results_Jpsi} compares the \mbox{HERA-B} and E866 measurements 
of the \jpsi polarization parameters, as functions of \xf and \pt,
with the corresponding curves computed with the model described in the previous section, 
using the central set of the CT14NLO~\cite{CT14NLO} proton PDFs,
properly adapting the calculations to the specific conditions 
(quarkonium state, collision energy, \pt and \xf coverage). 
The considered parameters are the \lth in the CS frame 
(where the model does not foresee visible deviations of \lph from zero, 
a prediction confirmed by the data, as seen in Fig.~\ref{fig:hierarchy}) 
and both \lth and \lph in the HX frame, 
where the two parameters share the magnitude of the natural polarization effect.
For each of the three scenarios, a range of values for the only parameter not fixed by our hypotheses, 
the \qqbar over $gg$ cross section ratio, $r$, 
has been determined so as to maximize the agreement with the data within the $\xf \lesssim 0.5$ domain:
the ranges are 4--5, 7--9 and 8--12, for the upper, central and lower scenarios, respectively.

\begin{figure}[t]
\centering
\includegraphics[width=0.75\linewidth]{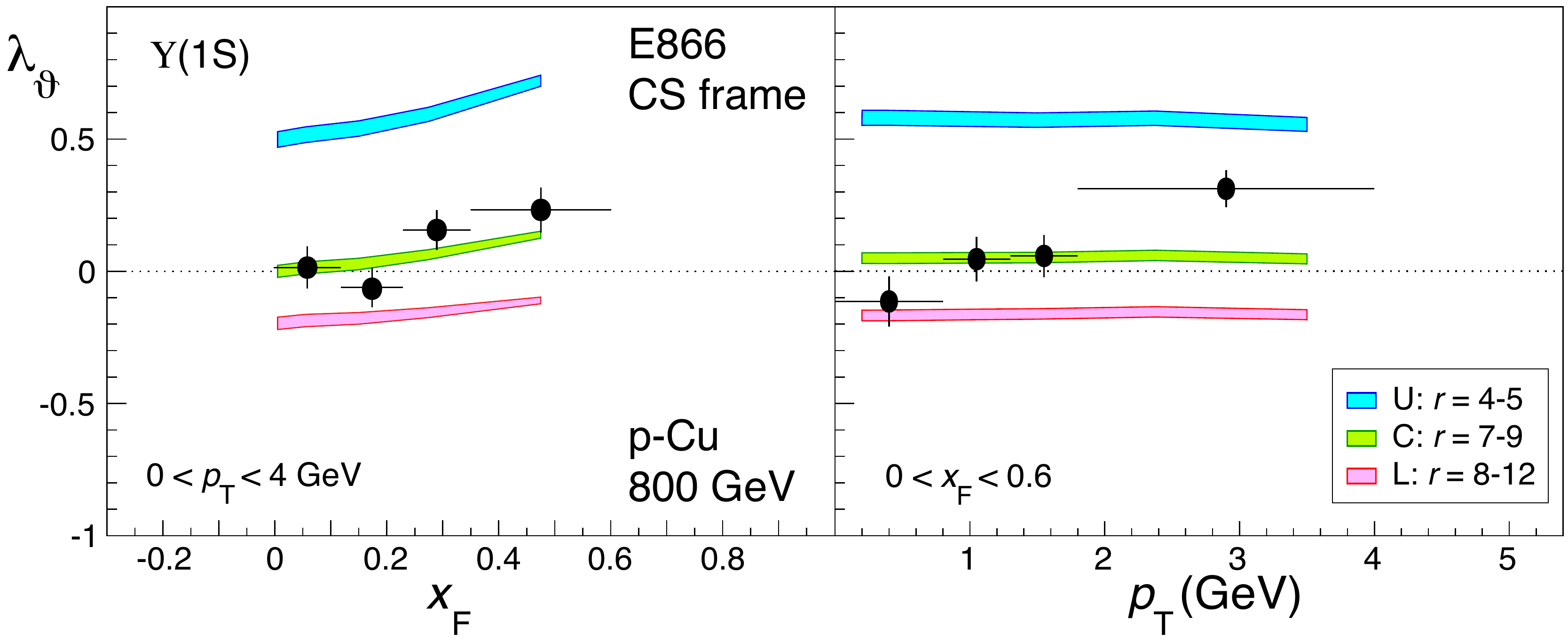}
\caption{Same as Fig.~\ref{fig:results_Jpsi}, 
for the $\Upsilon$(1S) \lth parameters, 
as measured by E866, in the CS frame.}
\label{fig:results_Upsilon1S}
\end{figure}

The uncertainty in the $\chi_c$ feed-down contribution 
has a large effect on the numerical determination of $r$, 
but almost no influence on the agreement between model and data. 
The interesting outcome of this comparison is that it is \emph{possible} 
to describe quite accurately the \jpsi data for $\xf \lesssim 0.5$, 
with the only substantial hypothesis that the directly produced vector quarkonium 
is transversely polarized along the relative direction of the colliding $q$ and $\overline{q}$, 
and longitudinally polarized along that of the colliding gluons.
This conclusion is reinforced by the comparison with the $\Upsilon$(1S) E866 measurement (\lth in the CS frame), 
shown in Fig.~\ref{fig:results_Upsilon1S} for the same $r$ values as determined using the \jpsi data. 
The central scenario is in very good agreement with the data. 
It is true that the fourth \pt point departs from the band, by around three times its uncertainty, 
but the significance of this difference seems to be suspiciously overestimated 
when we consider that the \lth value measured for \pt values only around 1\,GeV lower is perfectly reproduced by the model, 
and that almost no physical variations should be expected within such a small \pt interval, 
only one tenth of the particle mass.

The comparison considered until here involves 24 \jpsi and $\Upsilon$(1S) data points 
measured as a function of \xf, plus 29 points vs.\ \pt. 
Only a couple of points have central values differing from the central-scenario curves by around two or three times their uncertainties. 
While it is true that not all of these measurements are statistically independent, this remains a remarkable outcome, 
especially given the simplicity of the model and the diversity of measured patterns.
Therefore, we can state that the assumed interplay between \qqbar and $gg$ production, 
with their maximally different polarizations, 
describes the measured \jpsi and $\Upsilon$(1S) polarizations, 
within a reasonable set of assumptions for the unknown $\chi$ polarizations: 
all four states, $\chi_{c1}$, $\chi_{c2}$, $\chi_{b1}$ and $\chi_{b2}$, 
have angular momentum projections $\pm 1$ and 0, 
respectively along the $q$--$\overline{q}$ and $g$--$g$ collision directions, 
just as assumed for the directly produced vector mesons.
We remind that the $\chi$ feed-down fractions in the central scenario 
are fixed to the values 19\% and 45\%, respectively for the \jpsi and the $\Upsilon$(1S), 
where the latter is only a reasonable guess, given the absence of suitable measurements. 
For this central scenario, the ``universal'' ratio between the \qqbar and $gg$ cross sections 
for quarkonium production is determined to be $r = 8 \pm 1$.
The spread between the three bands (scenarios U, C and L) is much larger 
in Fig.~\ref{fig:results_Upsilon1S} than in Fig.~\ref{fig:results_Jpsi} because of the 
very uncertain $\chi_{b}$ feed-down fractions, mentioned before.

\begin{figure}[t]
\centering
\includegraphics[width=0.75\linewidth]{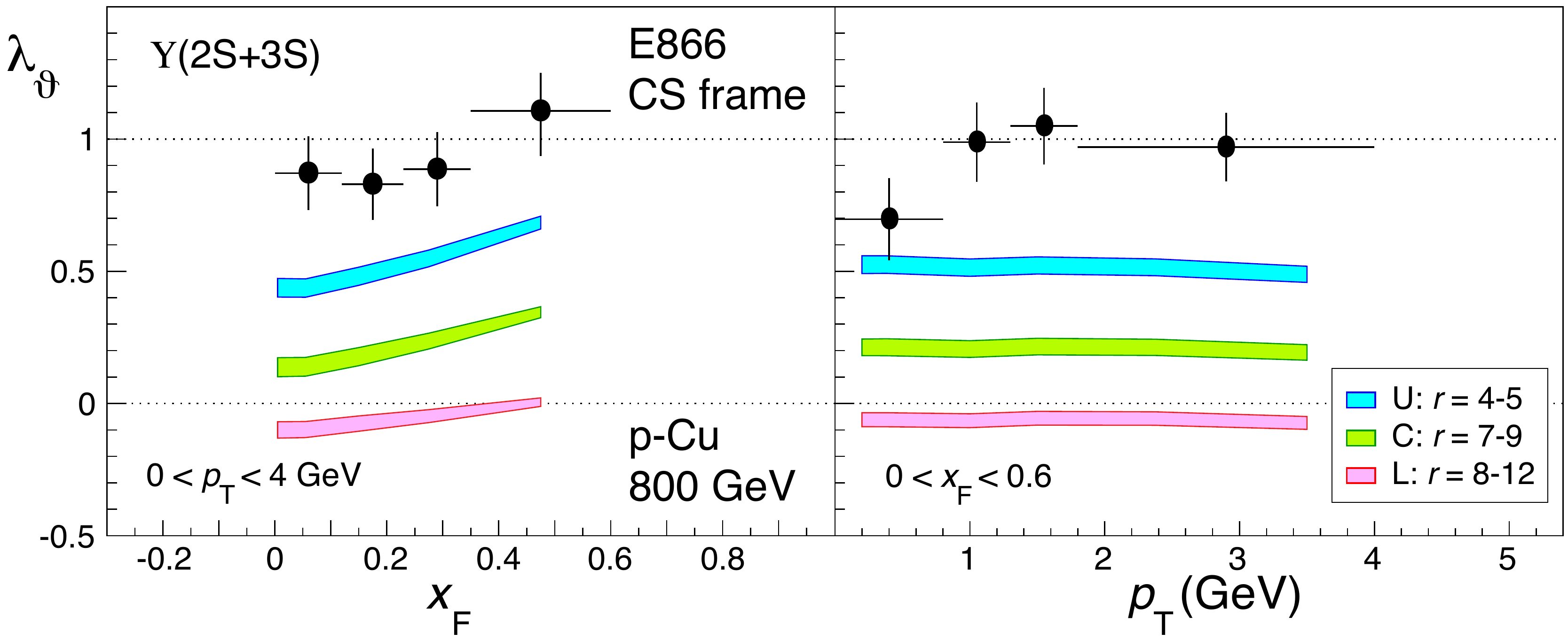}
\caption{Same as Fig.~\ref{fig:results_Upsilon1S}, 
for the $\Upsilon$(2S+3S).}
\label{fig:results_Upsilon23S}
\end{figure}

The last piece of our comparison concerns the E866 measurement for the
(unresolved) $\Upsilon$(2S) and $\Upsilon$(3S) states, 
shown in Fig.~\ref{fig:results_Upsilon23S}: 
the data are significantly above all three model scenarios. 
This would seem to imply that the (slightly) heavier $\Upsilon$ states 
evade the ``universality'' of the assumed physical inputs in a rather drastic way. 
In fact, even if we assumed that the polarizations of the $\chi_{b}$(2P) and $\chi_{b}$(3P)
states, contributing to the production of all three $\Upsilon$(nS) states, 
are extremely different from the polarizations of the $\chi_{b}$(1P) states, only
contributing to $\Upsilon$(1S) production (certainly a rather unreasonable hypothesis),
it would still be impossible to reproduce the data with our universal model for vector-quarkonium production: 
the only possibility would be to assume (possibly in addition) 
that the \qqbar to $gg$ cross section ratio is significantly higher for the only slightly heavier 2S and 3S states 
with respect to the 1S case, a hypothesis that would lead to a shift of all three bands towards the $\lth=\pm 1$ limit.

However, as already mentioned before, these $\Upsilon$(2S+3S) measurements 
are clearly a notable exception in the global panorama of the existing data. 
The first idea that comes to mind when seeing such a large discrepancy
is to look for something that makes the $\Upsilon$(2S+3S) measurement
significantly different from the \jpsi and $\Upsilon$(1S) measurements.
After all, polarizations are notoriously difficult to measure and 
it could well be that the reported values 
are affected by experimental challenges.
A first observation that can be made along those lines 
is that the \jpsi and $\Upsilon$(1S) states are seen as very prominent peaks
standing out of the underlying dimuon mass continuum,
while the $\Upsilon$(2S+3S) joint signal is not visible as a peak 
in the measured mass distribution (Fig.~1 of Ref.~\cite{E866-Upsilon}),
given the resolution and the signal-to-background ratio of the measurement.
A second interesting observation
is provided by the $\cos \vartheta$ distribution shown in Fig.~3-bottom of the E866 
publication\,\footnote{More precisely, in the figure of the arXiv version of the paper, 
given that the figure in the journal publication mistakenly shows the same distribution 
in the $\Upsilon$(1S) and $\Upsilon$(2S+3S) panels.}.
The curve displayed in that figure, which corresponds to the fourth \pt bin,
is distinctively asymmetric, 
thereby not corresponding to its legend 
(``$1 + 0.98 \cos^2 \vartheta$") and, more importantly, 
clearly departing from the parity-conserving $1 + \lth \cos^2 \vartheta$ shape 
that must apply to the dimuons produced in decays of quakonium states.
We have fitted either the negative or the positive hemispheres of the 
reported $\cos \vartheta$ distribution and obtained \lth values that differ
from each other by more than two times the published \emph{total} uncertainties,
an indication that those uncertainties might be underestimated.
Nevertheless, the large discrepancies seen in Fig.~\ref{fig:results_Upsilon23S} 
between the data points and the curves representing the central scenario
would not disappear even if the $\Upsilon$(2S+3S) \lth uncertainties 
would be increased by a factor of three.
It would be very valuable to redo the analysis of the E866 data, this time using 
the $\Upsilon$(2S+3S) over $\Upsilon$(1S) ratio as a function of $\cos \vartheta$,
which directly provides a measurement of the \emph{difference} between the 
two polarizations, with smaller systematic uncertainties thanks to the cancellation
of many potential effects in the ratio~\cite{CMS_chic_pol,CPT}.
Unfortunately, such a re-analysis is seemingly not possible~\cite{MikeLeitch},
so that we will need to wait for future measurements to fully clarify this puzzle.

\section{Data vs.\ model for pion-nucleus collisions}
\label{sec:pions}

Using the $r$ values determined from the proton-nucleus data 
(and the same three \jpsi feed-down scenarios in which the corresponding $r$ ranges were determined), 
we will now see how the model compares with the \jpsi polarization measurements performed with
pion beams.

\begin{figure}[t]
\centering
\includegraphics[width=0.5\linewidth]{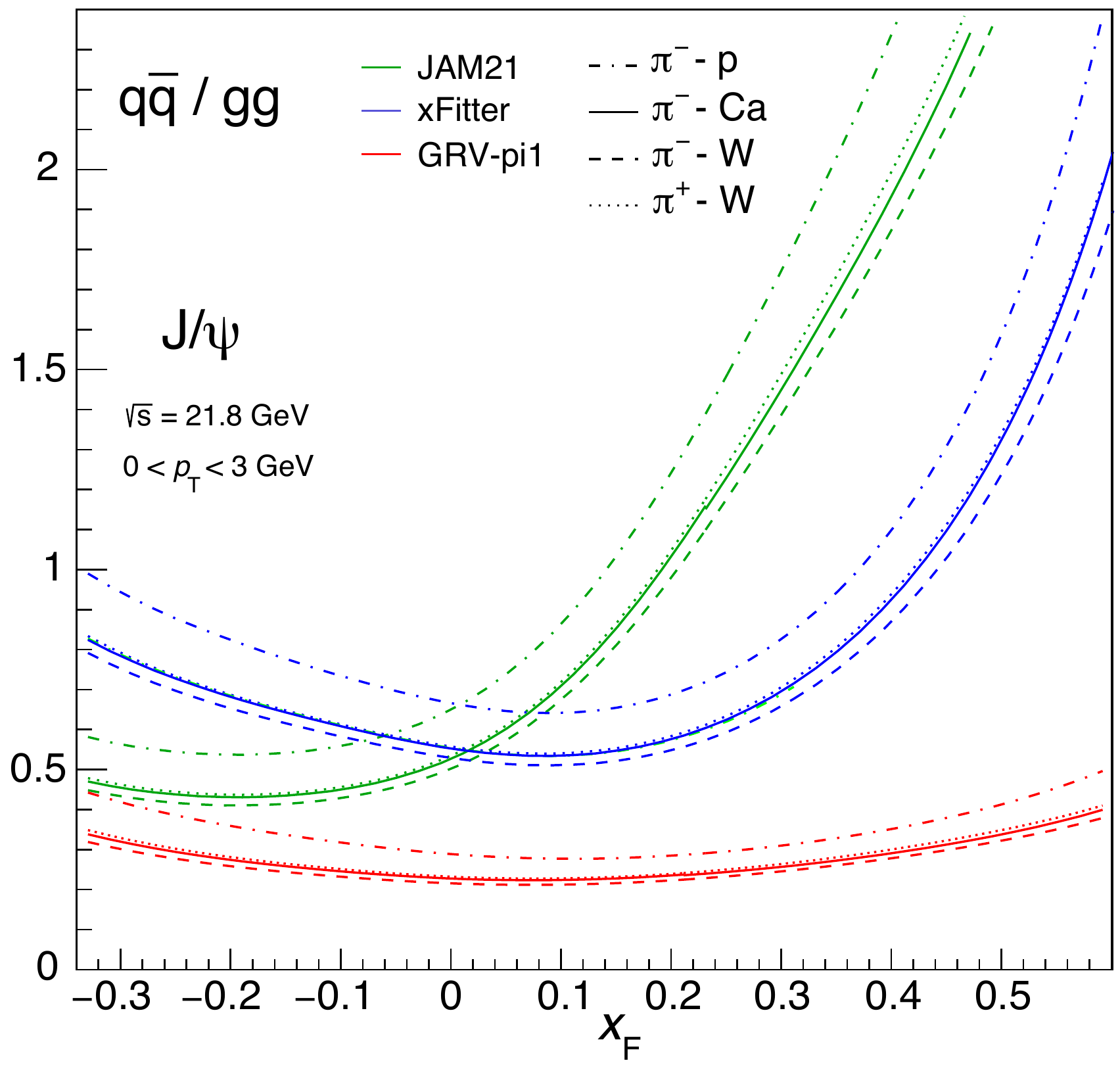}
\caption{The \qqbar over $gg$ parton luminosity ratios, vs.\ \xf,
for \jpsi production at the collision energy of E615~\cite{E615},
for the JAM21, xFitter, and \mbox{GRV-pi1} pion PDF sets,
and for the $\pi^-$p, $\pi^-$-Ca, $\pi^-$-W, and \mbox{$\pi^+$-W} collision systems.}
\label{fig:PDFratios_pions}
\end{figure}

Figure~\ref{fig:PDFratios_pions} shows the \qqbar over $gg$ parton luminosity ratios for \jpsi production, 
as computed using three different pion PDF sets: 
JAM21~\cite{JAM21}, xFitter~\cite{xFitter}, and \mbox{GRV-pi1}~\cite{GRVPI},
obtained through the LHAPDF package~\cite{LHAPDF}.
The \xf dependence of this ratio, as well as its average value and the covered \xf range, 
depend quite significantly on the chosen PDF set,
probably because of the poorly-known gluon density in the pion~\cite{Platchkov1,Platchkov2}. 
In comparison, the differences between positive and negative pion beams are negligible.
Also the nuclear effects, computed using the EPPS16 model~\cite{EPPS16}, 
have a minor impact, when we change the target from Ca to W, for example.

\begin{figure}[t!]
\centering
\includegraphics[width=0.7\linewidth]{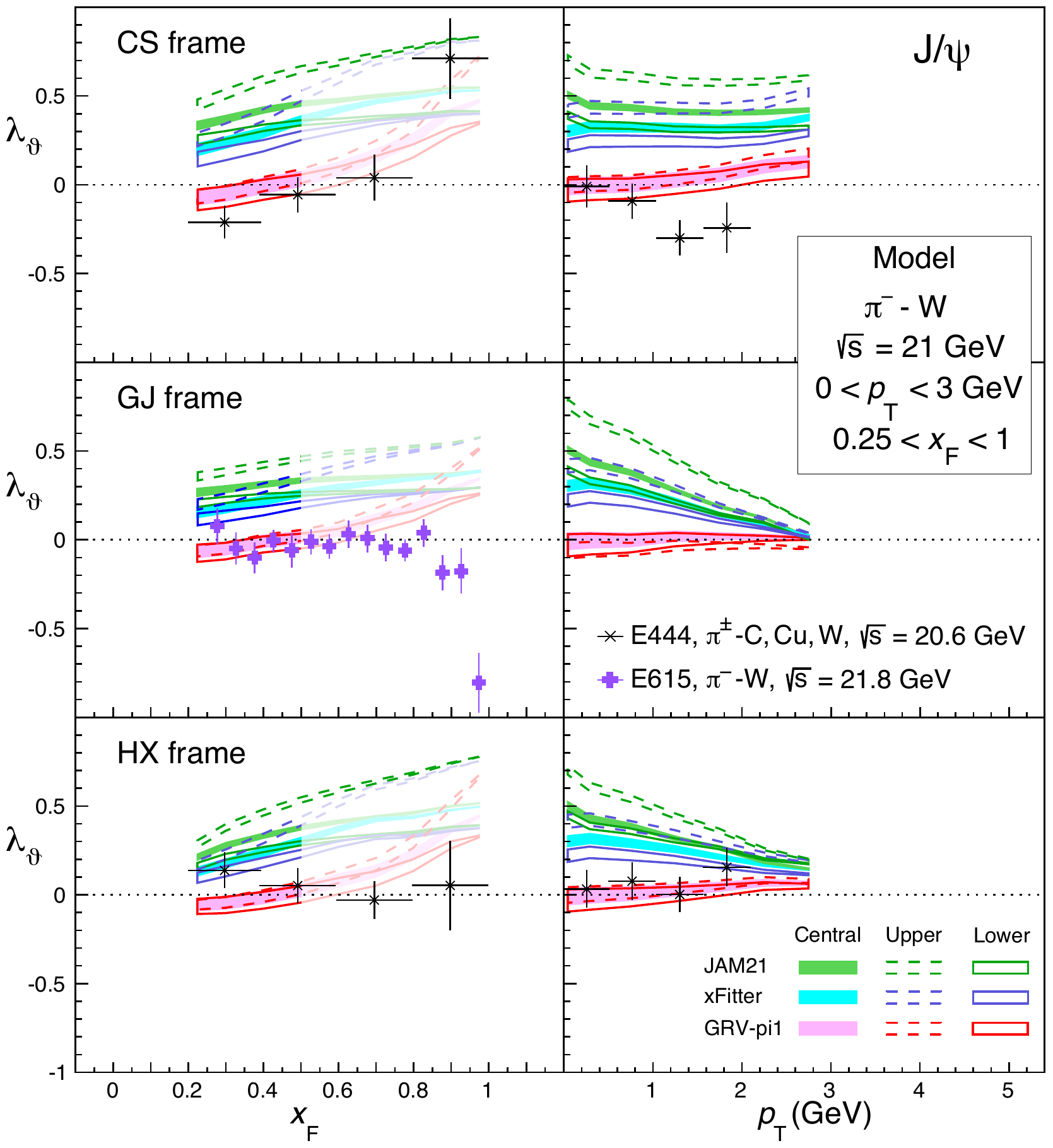}
\caption{The \xf and \pt dependences of the \jpsi polarization parameter \lth, 
in the CS, GJ and HX frames, as measured by E444 and E615 in $\pi$-W collisions.
The model predictions are computed using the
JAM21, xFitter and \mbox{GRV-pi1} pion PDF sets,
for the three feed-down scenarios.}
\label{fig:pion_results_21GeV}
\vglue5mm
\centering
\includegraphics[width=0.7\linewidth]{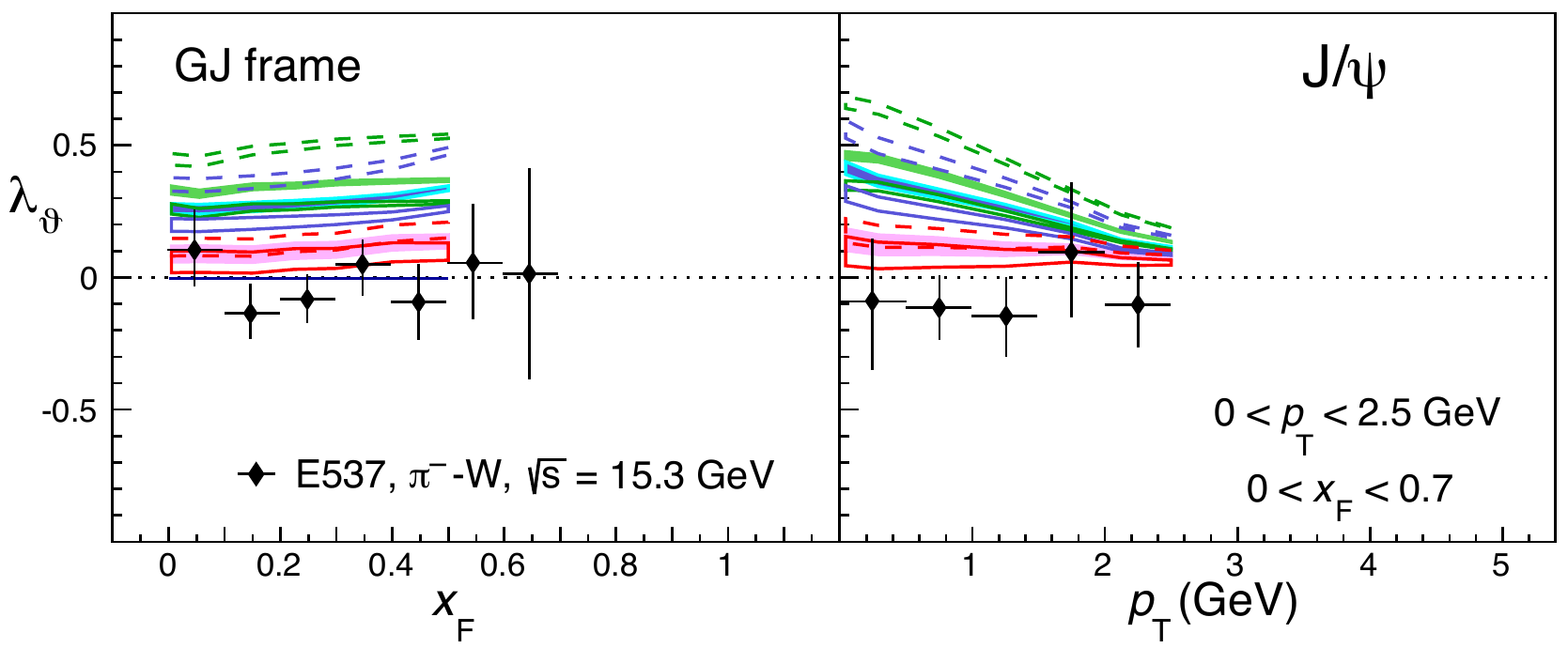}
\caption{Same as Fig.~\ref{fig:pion_results_21GeV}, for the E537 conditions.}
\label{fig:pion_results_E537}
\end{figure}

The dependence of the $\qqbar / gg$ ratio on the assumed PDF set 
directly translates into a corresponding variation in the polarization prediction,
as seen in Fig.~\ref{fig:pion_results_21GeV},
where the several \lth vs.\ \xf and \pt bands are computed for 
\mbox{$\pi^-$-W} collisions at $\sqrt{s} = 21$\,GeV, using the 
JAM21, xFitter and \mbox{GRV-pi1} pion PDF sets.
The filled bands represent the central scenario while the open ones represent the
lower (solid lines) and upper (dashed lines) cases.
We see that the E444 and E615 \lth measurements 
agree well with the model when the \mbox{GRV-pi1} parton densities are used,
and significantly depart from the bands representing 
the other two PDF sets.
Figure~\ref{fig:pion_results_E537} completes the data to model comparison, 
for pion-induced collisions, by showing the corresponding results for the E537 conditions,
characterized by a lower collision energy, $\sqrt{s} = 15.3$\,GeV, 
and a more backward \xf range, from 0 to 0.7.

We clearly see that accurate polarization measurements 
can provide precise constraints on global fits of pion PDFs, 
within the context of the rather general model discussed in this article
(which, as previously mentioned, might only be valid in the $\xf \lesssim 0.5$ range).

\section{Predictions for future measurements}
\label{sec:predictions}

\begin{figure}[t]
\centering
\includegraphics[width=0.55\linewidth]{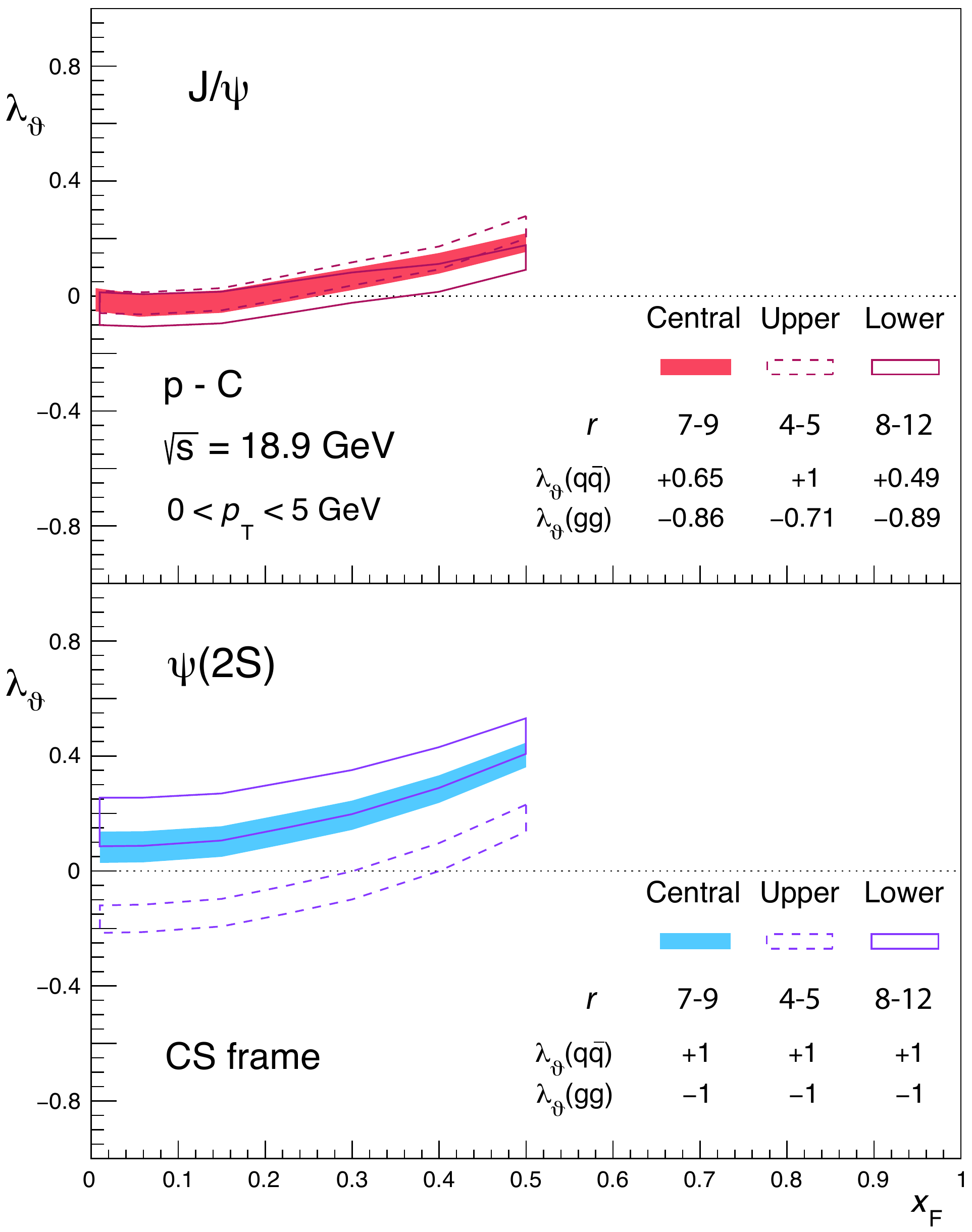}
\caption{The \xf dependence of the \lth polarization parameter, 
in the CS frame, as predicted for \jpsi (top) and $\psi$(2S) (bottom) production 
in p-C collisions at $\sqrt{s} = 18.9$\,GeV 
(corresponding to the conditions of the AMBER experiment), 
for the three considered \jpsi feed-down scenarios. }
\label{fig:results_AMBER_proton}
\end{figure}

New measurements in \emph{proton}-induced collisions can strengthen our confidence in the basic model,
where the observed polarization simply results from the interplay between \qqbar and $gg$ processes, 
and better tune its parameters.
After this further step of model validation, 
improved \emph{pion}-nucleus data can effectively represent a sensitive constraint on the pion PDFs.
The AMBER experiment~\cite{AMBER} at the CERN SPS accelerator, for example,
is expected to collect large samples of \jpsi events using 190\,GeV proton and pion beams, 
with carbon and tungsten targets. 
The top panel of Fig.~\ref{fig:results_AMBER_proton} shows, for p-C collisions in the conditions of AMBER,
the predicted \xf dependence of the \jpsi \lth parameter, in the CS frame, 
for the same three feed-down scenarios considered above. 
We do not show the \lph observable because, in this frame, 
it is perfectly compatible with being zero and flat with \pt; correspondingly, 
the frame-independent parameter \ltilde~\cite{Faccioli:2010ej,bib:Faccioli-PRD-FrameInv}
is essentially indistinguishable from \lth (and is, hence, also not shown).

The corresponding predictions for $\psi$(2S) production are presented in the bottom panel.
This measurement would be particularly interesting because it would test the model 
and constrain the pion PDFs in a cleaner and stronger way, 
free from the $\chi_c$ feed-down uncertainties: 
in this case, we simply have $\lth = -1$ and $+1$ for $gg$ fusion and \qqbar annihilation, respectively, 
as for \emph{direct} \jpsi production. 
The $r$ range for $\psi$(2S) production is assumed to be the same as for the \jpsi, 
indirectly depending, therefore, on the \jpsi feed-down scenario.
This is why, in the figure, three sets of $\psi$(2S) predictions are shown, 
despite the fact that $\psi$(2S) production is intrinsically independent of feed-down.

The assumed equality of $r$ for \jpsi and $\psi$(2S) production 
follows the spirit of the factorization hypothesis motivating both the CEM and NRQCD models: 
we should, in fact, expect a cancellation of the long-distance bound-state formation effects, 
possibly differentiating $\psi$(2S) and \jpsi, in the ratio of the $\qqbar / gg$ partonic cross sections 
(Eq.~\ref{eq:R_parton_xsections}).

The comparison between the \jpsi and $\psi$(2S) predictions shows that 
simultaneous measurements of both polarizations in proton-nucleus collisions 
can be used to fix the $r$ range \emph{and} determine the best \jpsi feed-down scenario, 
at the same time.

\begin{figure}[t]
\centering
\includegraphics[width=0.55\linewidth]{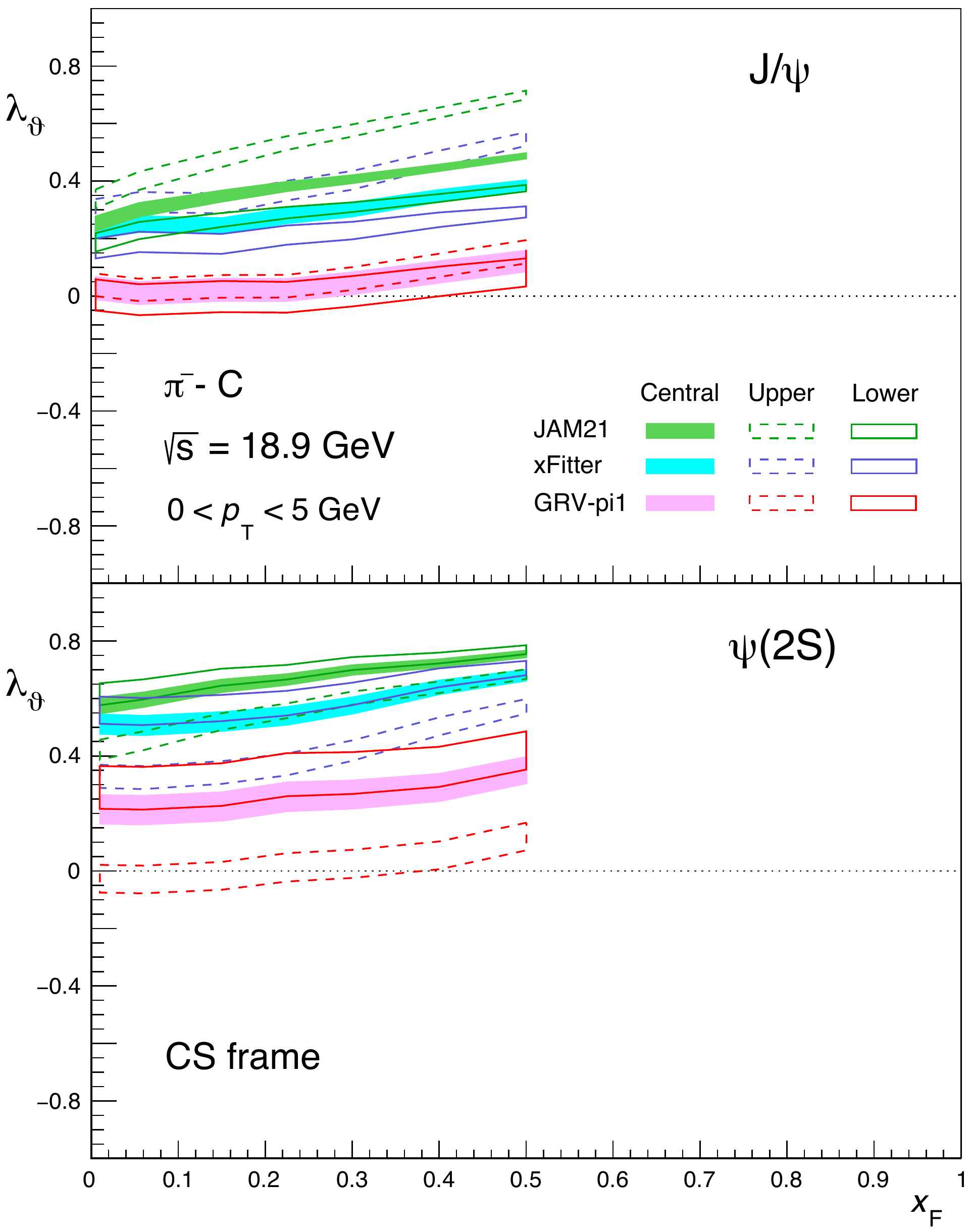}
\caption{Same as Fig.~\ref{fig:results_AMBER_proton}, for $\pi^-$-C collisions. 
The bands represent the model predictions obtained in the three polarization scenarios, 
using the three pion PDF sets.}
\label{fig:results_AMBER_pion}
\end{figure}

The corresponding \jpsi and $\psi$(2S) predictions for pion-induced collisions 
are shown in Fig.~\ref{fig:results_AMBER_pion}. 
Here, the attention should be focused on the difference between the results obtained 
with the three PDF sets: the spread of polarization values, on average of order $\Delta \lth \simeq 0.3$, 
shows that the measurement should be able to significantly discriminate between existing pion PDF sets
or, alternatively, provide strong constraints on the future determination of new PDF sets.

\section{Summary}
\label{sec:summary}

We proposed an interpretation of existing low-\pt quarkonium polarization measurements,
performed in both proton-nucleus and pion-nucleus collisions,
on the basis of a single and simple hypothesis: 
the production is dominated by 2-to-1 \qqbar and $gg$ processes, 
leading, respectively, to transversely and longitudinally polarized S-wave quarkonia, when directly produced. 
The feed-down from $\chi$ states is modelled through three hypotheses, reflecting the existing data,
the central one being complemented by two extreme scenarios that, taken together, 
can be seen as representing the uncertainty that this input has on the results.

The \jpsi and $\Upsilon$(1S) data are well reproduced by the model for $\xf \lesssim 0.5$;
the trend to longitudinal polarization seen towards the high-\xf edge is
a long-debated phenomenon that is not addressed in the present study. 
We have been unable to reproduce, 
even invoking barely reasonable variations in the feed-down contributions, 
the very large difference observed by E866 between the $\Upsilon$(2S+3S) and $\Upsilon$(1S) polarizations: 
an experimental problem, 
possibly related to the challenging discrimination of the $\Upsilon$(2S+3S) signal from the background, 
might be the most reasonable explanation for this puzzling observation.
More generally, some of the publications have not reported a complete evaluation of the uncertainties
affecting the (sometimes rather old) measurements, 
which might explain the presence of suspicious fluctuations.
It is clear that improved measurements are needed
for a better understanding of the picture of quarkonium production at low \pt.

The calculations made for the comparisons with the pion-nucleus data 
reveal an interesting correlation between the \jpsi polarization and the assumed parton densities:
the predicted polarization depends on the adopted PDF set much more than on any other model input, 
directly reflecting the large variation of the $\qqbar / gg$ parton luminosity ratio, 
for which the three considered sets differ by a factor of 2 or more, with a strong dependence on \xf and \pt. 
While the GRV set reproduces reasonably well the existing data, 
the uncertainty represented by the differences between the three sets clearly indicates that the polarization observable 
has the potential to provide a strong constraint on the pion PDFs.
With this motivation, we reported predictions for AMBER, 
as an example of an experiment capable of performing a further validation of our simple model, 
through improved \jpsi (``calibration'') measurements in proton-nucleus collisions, and, subsequently, 
constrain the pion PDFs with corresponding measurements in pion-nucleus collisions.
We also emphasized the important role that a measurement of the $\psi$(2S) polarization can have, 
by virtue of its independence from hypotheses on the $\chi_c$ feed-down contributions, 
both to validate the model and to determine the pion PDFs.

\bigskip
P.F.\ and C.L.\ acknowledge support from 
Funda\c{c}\~ao para a Ci\^encia e a Tecnologia, Portugal, 
under contract CERN/FIS-PAR/0010/2019

\bibliographystyle{cl_unsrt}
\bibliography{low_pT_quarkonium-pol}{}

\end{document}